\documentclass[prl,twocolumn,showpacs,preprintnumbers,amsmath,amssymb, superscriptaddress]{revtex4-1}

\usepackage[makeroom]{cancel}

\usepackage{amsmath}    
\usepackage{amssymb}
\usepackage{graphicx}   
\usepackage{verbatim}   
\usepackage{color}      
\usepackage{hyperref}   
\usepackage[normalem]{ulem}
\usepackage{natbib}
\usepackage{fixmath}
\usepackage{enumitem}

\hypersetup{colorlinks,linkcolor=blue,urlcolor=blue,citecolor=blue}
\newcommand{\beq}{\begin{equation}}
\newcommand{\eeq}{\end{equation}}
\newcommand{\bea}{\begin{eqnarray}}
\newcommand{\eea}{\end{eqnarray}}

\newcommand{\be}{\begin{equation}}
\newcommand{\ee}{\end{equation}}

\definecolor{darkgreen}{rgb}{0,0.5,0}
\definecolor{orange}{rgb}{1,0.5,0}
\definecolor{grey}{rgb}{.6,.6,.6}

\newcommand{\bra}[1]{\langle #1|}
\newcommand{\ket}[1]{|#1\rangle}
\newcommand{\average}[1]{\langle #1\rangle}

\newcommand{\imp}{{\rm imp}}

\newcommand{\bS}{{\vec S}}
\newcommand{\bk}{\mathbf k}

\newcommand{\br}{\mathbf r}

\newcommand{\bs}{\vec s\;}
\newcommand{\bsigma}{{{\vec \sigma}\;}}

\newcommand{\rmd}{\text{d}}

\begin{document}
\title{The fate of the Kondo cloud in a superconductor}

\author{C\u at\u alin Pa\c scu Moca}
\affiliation{MTA-BME Quantum Dynamics and Correlations Research Group, E\"otv\"os Lor\'and Research Network} 
\affiliation{Department of Physics, University of Oradea, 410087, Oradea, Romania}
\author{Ireneusz Weymann}
\affiliation{Institute of Spintronics and Quantum Information,
	Faculty of Physics, Adam Mickiewicz University, 61-614 Pozna\'n, Poland}
\author{Mikl\'os Antal Werner}
\affiliation{MTA-BME Quantum Dynamics and Correlations Research Group, E\"otv\"os Lor\'and Research Network} 
\affiliation{Department of  Theoretical Physics, Budapest University of Technology and Economics, 
Budafoki \'ut 8., H-1111 Budapest, Hungary}
\author{Gergely Zar\'and}
\affiliation{MTA-BME Quantum Dynamics and Correlations Research Group, E\"otv\"os Lor\'and Research Network} 
\affiliation{BME-MTA Exotic Quantum Phases 'Lend\"ulet' Research Group, Institute of Physics, Budapest University of Technology and Economics, 
Budafoki \'ut 8., H-1111 Budapest, Hungary}

\date{\today}

\begin{abstract}
Magnetic impurities embedded in a metal are screened by the Kondo effect, signaled by the formation of an extended
 correlation cloud, the so-called Kondo or screening cloud. In a superconductor,  
the Kondo state turns into  sub-gap Yu-Shiba-Rusinov (Shiba) states, 
and a quantum phase transition occurs between  screened and unscreened   phases    
once the superconducting energy gap $\Delta$ becomes sufficiently large compared to   the Kondo temperature, $T_K$.
Here we show that, although the Kondo state does not form in the unscreened phase, 
the Kondo cloud  does exist in both quantum phases.
However,  while  screening is complete  in the screened phase,  it is only partial in the unscreened phase.
Compensation, a quantity introduced to characterize the integrity of the cloud, is  universal,   
and shown to be related to  the magnetic impurities'  $g$-factor,  monitored experimentally by bias spectroscopy.
\end{abstract}

\maketitle

\paragraph{Introduction.---} One of the most fascinating manifestations of magnetic interactions in metals is 
the Kondo effect~\cite{Hewson.1992}, where a local spin interacts with a sea of non-interacting electrons, 
to get there completely dissolved by quantum fluctuations below the so-called  
Kondo temperature, $T_K$.  This magic quantum spin vanish is accompanied by 
the formation of the so-called \emph{Kondo cloud},  as characterized by the ground state correlation function
\begin{equation}C(\br) \equiv \average{\bS_\imp\, \cdot \bs(\br)}\,,
\label{eq:chi}
\end{equation}
with $ \bs(\br)$ 
the electrons' spin density at position $\br$, and $\bS_\imp$ the 
spin of the magnetic impurity, which we assume to be of size $S_\imp = 1/2$, 
which is typical in quantum dot devices.
The antiferromagnetic correlations in Eq.~\eqref{eq:chi} have been investigated 
theoretically~\cite{Scalapino.1987,Chen.1987,Chen.1992, Affleck.1996a, Affleck.1996b,Affleck.2001,Costamagna.2006,Hand.2006, Borda.2007,Affleck.2008,Bergmann.2008, Holzner.2009, Dagotto.2010,Mitchell.2011,Medvedyeva.2013,Lechtenberg.2014,Ghosh.2014,Arrigoni.2015, Florens.2015}  and also attempted to be measured experimentally  by many~ \cite{Boyce.1974, Pruser.2011,Jiang.2011, Figgins.2019}. 
They oscillate fast in space,  and are characterized by an exponentially large length scale, 
the so-called Kondo  scale, $\xi_K \approx v_F/T_K$, with $v_F$ the Fermi 
velocity~\footnote{We use units  $\hbar = k_B = 1$.}.    In $D$ spatial dimensions,--  apart from logarithmic corrections~\cite{Borda.2007, Affleck.2008,Affleck.2010},-- the 
envelope of $C(r)$  decays as $\sim 1/r^D$ at short distances, $r< \xi_K$, 
while it falls off as $\sim 1/r^{D+1}$ for $r \gg \xi_K$.
Simple estimates yield the Kondo scale  $\xi_K$  as large as $\sim 1\mu \rm m$
in typical metals, a distance comparable to the physical dimensions of mesoscopic devices.

The antiferromagnetic correlations residing in this huge Kondo cloud are, however, quite small,
as signaled by the sum rule~\cite{Borda.2007,Borda.2009}
\begin{equation}
\int   \; \average{\bS_\imp\, \cdot \bs(\br)} \; {\rm d}^D\br =  -\frac {3}{4}\;\kappa\;\;,
\label{eq:integral}
\end{equation}
with $ \average{\dots}$ referring to the ground state average, and $\kappa=1$ a certain
measure of  quantum screening, introduced later. Equation~\eqref{eq:integral} just expresses 
that, after all, there is only a \emph{single}  spin that is needed to forms a singlet state
with the impurity, and that this compensating conduction electron spin is smeared in the Kondo volume,  $\sim \xi_K^D$.
Entanglement entropy~\cite{Bayat.2010,Bayat.2012} calculations and the study of  entanglement witness operators~\cite{Lee.2015} 
also corroborate this picture, and confirm that the local spin's entanglement, i.e. the Kondo cloud resides within a distance $\xi_K$ from 
the impurity. Although many theoretical proposals have been put forward to measure the Kondo cloud 
by now~\cite{Affleck.2001,Affleck.2008,Yuval.2013},
the cloud remained elusive for experimentalists for a very long time~\cite{Boyce.1974,Pruser.2011,Jiang.2011,Figgins.2019}, and its 
large extension has only been confirmed very recently via  Fabry-P\' erot oscillations in a mesoscopic 
system~\cite{Tarucha.2020}.
\begin{figure}[b!]
	\includegraphics[width=0.85\columnwidth]{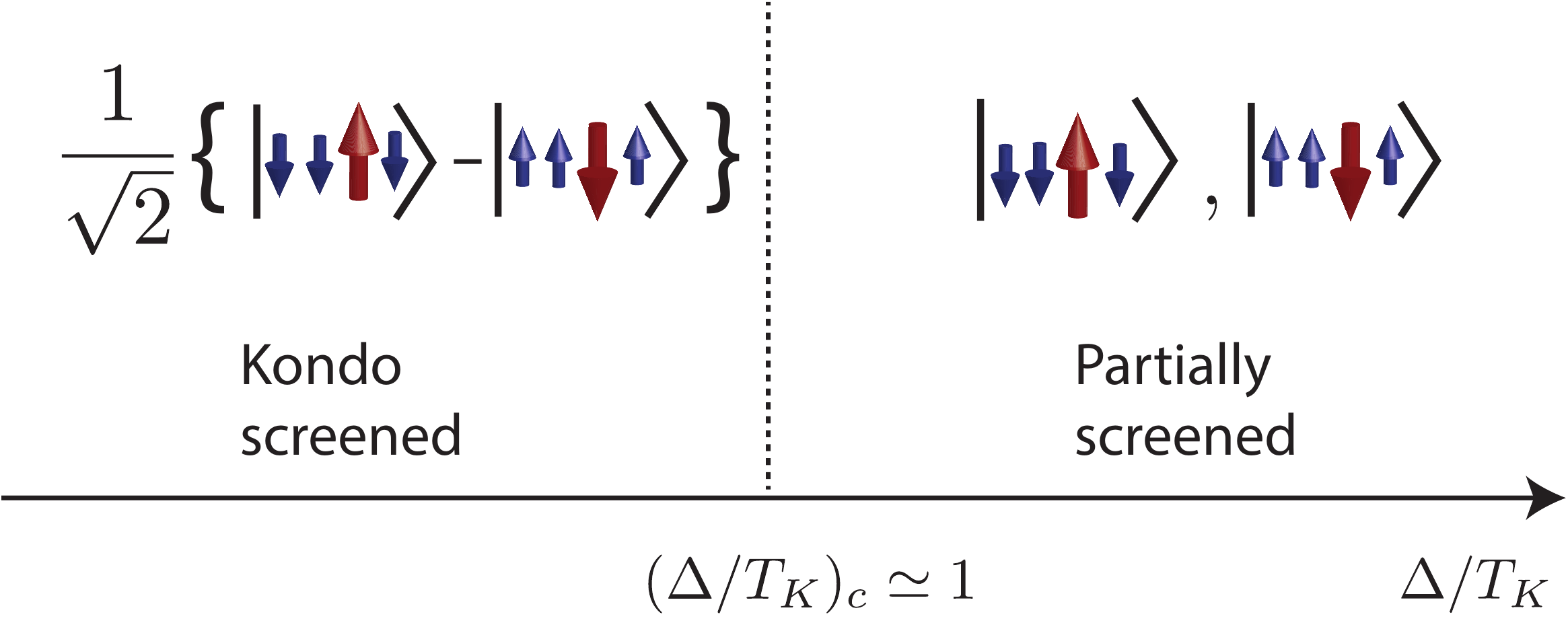}
	\caption{Schematic ”phase diagram” of the model at zero temperature. When $\Delta > T_K$, the ground state
	is a doublet with an asymptotically free spin decoupled from the superconductor, while in the opposite limit,
	when $\Delta<T_K$, the ground state is a many-body singlet.}
	\label{fig:phase_diagram}
\end{figure}
In this work we investigate the fate of the Kondo compensation cloud in an $s$-wave  superconductor. In a 
superconductor, the superconducting gap $\Delta$  competes with the Kondo effect, and prohibits screening 
of the magnetic impurity for weak interactions, $T_K\ll \Delta$. In this case, the magnetic impurity spin remains free even at very 
small temperatures, but it binds superconducting quasiparticles to itself antiferromagnetically, amounting to 
discrete (singlet) sub-gap electron and hole excitations~\footnote{These two Shiba states 
are not distinct, they are the electronic and hole parts of the same quantum state}, called the Shiba states~\cite{Yu.1965,Shiba.1968,Rusinov.1974}.
Beyond a critical magnetic coupling, i.e. for  $\Delta/T_K < (\Delta/T_K)_c\approx1.1$~\footnote{The critical value depends on the precise definition of the Kondo temperature, $T_K$. Throughout this work, we define $T_K$ as the half-width of the Kondo resonance,
	i.e., that of the composite fermion's spectral function~\cite{Costi.2000}.},
a first order quantum phase transition occurs, and the subgap singlet excitation becomes the ground state, 
as illustrated in Fig.~\ref{fig:phase_diagram}. A spin $S_\imp = 1/2$ impurity embedded
into a superconductor has therefore two quantum phases,
a screened \emph{singlet phase} for $ \Delta/T_K < (\Delta/T_K)_c $, and a \emph{doublet phase} 
for $\Delta/T_K > (\Delta/T_K)_c $.

Here we investigate the structure of the Kondo compensation cloud in these two phases. Somewhat surprisingly, we find that 
the superconductor does \emph{not} destroy the Kondo cloud even in the unscreened doublet quantum phase, just reduces 
the degree of compensation, $\kappa$, from its value $\kappa=1$ in the singlet phase to $\kappa=\kappa(\Delta/T_K)<1$
in the doublet quantum phase. 
We dub the corresponding  fractional compensation cloud as the \emph{Shiba cloud}. 
The fractional compensation emerges as a result of the competition of the Kondo screening length, $\xi_K$, and 
the superconducting correlation length, $\xi$, and in the doublet phase the extension of the cloud is just 
the coherence length, $\xi$, rather than $\xi_K$. This enormous extension of the Shiba cloud is in agreement 
with recent experiments on side-coupled superconducting quantum dot devices,
measuring the size of Shiba states~\cite{Scherubl.2020}.

\paragraph{Compensation.–-} 
We first show that Eq.~\eqref{eq:chi} is satisfied with $\kappa = 1$ in the singlet phase,  $\Delta/T_K  < (\Delta/T_K)_c$. To prove 
Eq.~\eqref{eq:chi},  we only need to exploit SU(2) symmetry and the fact that the ground state $|G\rangle$ is a singlet, implying that 
$|G\rangle$  is an eigenstate of the total spin operator,  $\bS_T$, with zero eigenvalue, 
$$
\bS_T |G\rangle = (\bS_\imp + \int {\rm d}^D\br\;  \bs(\br)) |G\rangle = 0\;. 
$$
Multiplying this equation by $\langle G|\,\bS\dots $ from the left and using  $\bS_\imp\cdot \bS_\imp = 3/4$ yields immediately 
Eq.~\eqref{eq:integral} with $\kappa=1$.  

We now show that  a similar relation holds even in the doublet phase, 
but with $\kappa<1$, defining the degree of \emph{compensation}. In the doublet phase, we 
have two degenerate ground states, $|\! \Uparrow\rangle$ and $ |\!\Downarrow\rangle$. 
These two states  transform among each other upon the action of the total spin operators as
$$
\bS_T|\alpha\rangle  = \sum_\beta \frac 1 2 \;{\vec \sigma}_{\beta\alpha} |\beta\rangle\,,
$$
with $\alpha$ and $\beta$ referring to  $|\!\Uparrow\rangle$ and $|\!\Downarrow\rangle$, and 
$\boldsymbol{\sigma}$  the  Pauli matrices.
 Similar to the spin $S_T = 0$ case, we now multiply this equation by  $\langle \alpha|\,{\vec S}_\imp\dots $, and average over 
$\alpha$. On the right hand side, however, we can now use the Wigner-Eckart theorem, according to which 
$$
\langle\alpha| {\vec S}_{\;\imp}|\beta\rangle = g \; \frac 1 2 \bsigma_{\alpha\beta}\;,
$$ 
with $g$ the $g$-factor of the impurity spin. This immediately yields Eq.~\eqref {eq:integral} with 
\begin{equation}
\kappa = 1-g\;.
\end{equation} 
For a free spin we have $g=1$, implying no compensation, $\kappa = 0$. However, as we discuss below, 
for a spin embedded into a superconductor, $g$ becomes finite due to quantum fluctuations, leading to a partial 
compensation of the spin and a squeezed Kondo cloud. 

\paragraph{Perturbation theory.---} 
In the limit $\Delta\gg T_k$ perturbation theory and a renormalization group approach  
can be used to assess the origin of $g$. We consider for that the Kondo model 
\begin{equation}
H = J\, \bS_{\rm imp}\cdot \bs(0) +H_{\rm host}\, ,
\label{eq:Hamiltonian}
\end{equation}
with  $J$  the  local Kondo coupling, and 
 $ \bs(0) = \frac 1 2 \psi^\dagger(0) \mathbold{\sigma} \psi(0) $ the spin density at the origin, 
expressed now in terms of the conduction electrons' field operator, $\psi_\sigma(\br) = \sum_{\bk} e^{i\bk \br}/\sqrt{V}\; c_{\bk\sigma}$. 
The term  $H_{\rm host}$  describes the non-interacting  superconducting host, 
 $$
 H_{\rm host} = \sum_{\bk,\sigma}  \epsilon_\bk \,c^\dagger_{\bk\sigma} c_{\bk\sigma} + 
 \sum_{\bk,\sigma}   (\Delta \,c^\dagger_{\bk\uparrow} c^\dagger_{-\bk\downarrow} + \text{h.c.})\;.
$$

To determine $\kappa$, we simply compute $\langle\Uparrow\! |  S_\imp^z| \!\! \Uparrow\rangle = g / 2$ 
perturbatively in $J$.  A straightforward calculation yields~\cite{SM}
\begin{equation}
\kappa = 1- g = \frac {j_0^2} 4 \ln\left(\frac {\Lambda_0} \Delta \right) + {\cal O}(j_0^3)\;,
\label{eq:kappa_perturbative}
\end{equation} 
with $\Lambda_0$ a bandwidth cutoff of the order of the Fermi energy, and $j_0 =J \varrho_0$ the usual dimensionless Kondo 
coupling, defined by means of the local density of states at the Fermi energy, $\varrho_0$. 
Clearly, the compensation contains a logarithmic singularity, which  must be handled by resumming  
the perturbation series up to infinite order. We have performed this resummation in subleading (so-called leading logarithmic)
order  by using the multiplicative renormalization group (RG)~\cite{SM}, and exploiting the invariance of 
the impurity contribution to the free energy under the RG. This calculation 
yields the expression
\begin{equation}
\kappa = 1- \exp\left[ \frac1 2 \,  {j_0}   - \frac 1 2 \, {j(\Delta/T_K)}   \right]\;,  
\label{eq:kappa}
\end{equation}
with $j(\Delta/T_K)$ the renormalized exchange coupling, 
\begin{equation}
j(\Delta/T_K)\approx \frac 1 {\ln\left( \frac {{\cal F} \Delta} {T_K}  \right) - \frac 1 2 \ln \left(\ln\left( \frac {{\cal F} \Delta} {T_K}  \right)  \right) }\;.
\label{eq:j(delta/T_K)}
\end{equation}
Here $T_K = \Lambda_0 \,{\cal F}\, \sqrt{j_0} \, e^{-1/j_0}$ denotes  the Kondo temperature in the next to leading logarithmic approximation,
with  ${\cal F}\approx 2.5$  determined numerically to fit the Kondo temperature, defined as the half-width of the Kondo resonance~\cite{Costi.2000}.  
Obviously, in the limit $j_0\to 0$, Eq.~\eqref{eq:kappa}   becomes  a universal function, $\kappa=\kappa(\Delta/T_K)$.

\paragraph{Numerics.---}
 
\begin{figure}[t!]
	\includegraphics[width=0.9\columnwidth]{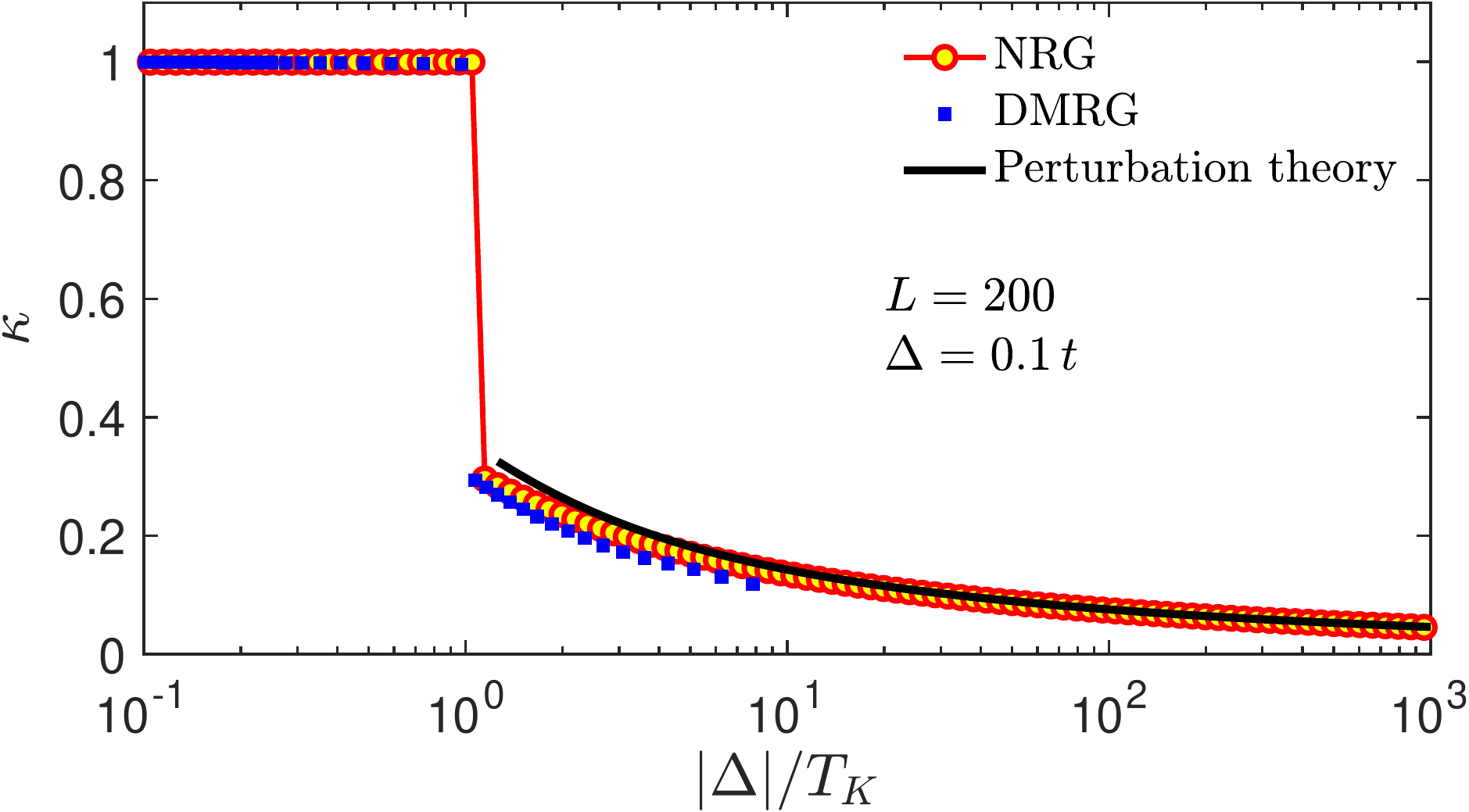}
	\caption{ Compensation $\kappa$ across the quantum phase transition as a 
	function of $\Delta/T_K$, for $j_0 = 0.05$. In the fully screened
	 regime, $\Delta/T_K\lesssim 1.1$,
	  $\kappa$ remains  $1$,  and it displays  a universal jump of size 
	$\Delta\kappa \simeq 0.719$ at the quantum phase transition,  followed by a monotonous decrease 
	 in the partially screened regime. Blue and yellow squares represent DMRG  
	and NRG results, while the solid line presents the theoretical result,   Eq.~\eqref{eq:kappa}.}
	\label{fig:kappa}
\end{figure}

To verify the above scenario and to determine the compensation 
$\kappa(\Delta/T_K)$ accurately, we carried out detailed numerical simulations using numerical renormalization group (NRG)~\cite{Wilson.1975} 
as well as density matrix renormalization group (DMRG)~\cite{White.1996, Schollwock.2005}
  methods. 
In both approaches, we can compute the ground state expectation value of the local spin, 
extract the $g$-factor from that,  and express the compensation $\kappa$ as
\begin{equation}
\kappa =1-2\bra{\Uparrow}S_{\rm imp}^{z} \ket{\Uparrow}
\label{eq:compensation}
\end{equation} 
in the unscreened phase. The results are  presented in Fig.~\ref{fig:kappa}.
They  show perfect agreement with each other, and also with the analytical expressions, 
Eqs.~\eqref{eq:kappa} and \eqref{eq:j(delta/T_K)}. The compensation right at the quantum phase transition 
is around $\kappa_c\approx 0.28$, thus quantum fluctuations screen around 1/3'd of the total spin, even in the 
doublet  phase.

The build-up of finite compensation is accompanied by the evolution of the screening cloud.
We can directly monitor this latter  in one dimension  with  DMRG computations.  
 In the absence of superconductivity,--  apart from an oscillating part $\sim \cos(2\, k_F x)$,– 
 spin-spin   correlations decays as 
$|C(x)| \sim \xi_K/x$ at short distances, $x\ll \xi_K$,  while they fall off quadratically for $x\gg \xi_K$, where 
 $|C(x)|\sim (\xi_K/x)^2$~\cite{Ishii.1978,Affleck.1996b, Affleck.2008, Borda.2009}. The power law decay originates in both regimes
from electron-hole excitations. In a superconductor, however, electron-hole excitations of energy $\delta E<2\Delta$ 
are forbidden. Correspondingly, the power law behavior is suppressed beyond the 
associated superconducting correlation length, $\xi = \Delta/v_F$, beyond which  correlations 
show an exponential decay, as also demonstrated by perturbation theory (see Ref.~\cite{SM}). 
The Shiba phase transition occurs right   when the Kondo and coherence 
lengths become approximately  equal, $\xi\approx \xi_K$: the spin becomes  fully screened under the condition that 
the Kondo compensation cloud  fits into the coherence volume $\sim   \xi^D$.

\begin{figure}[tbh!]
	\includegraphics[width=0.9\columnwidth]{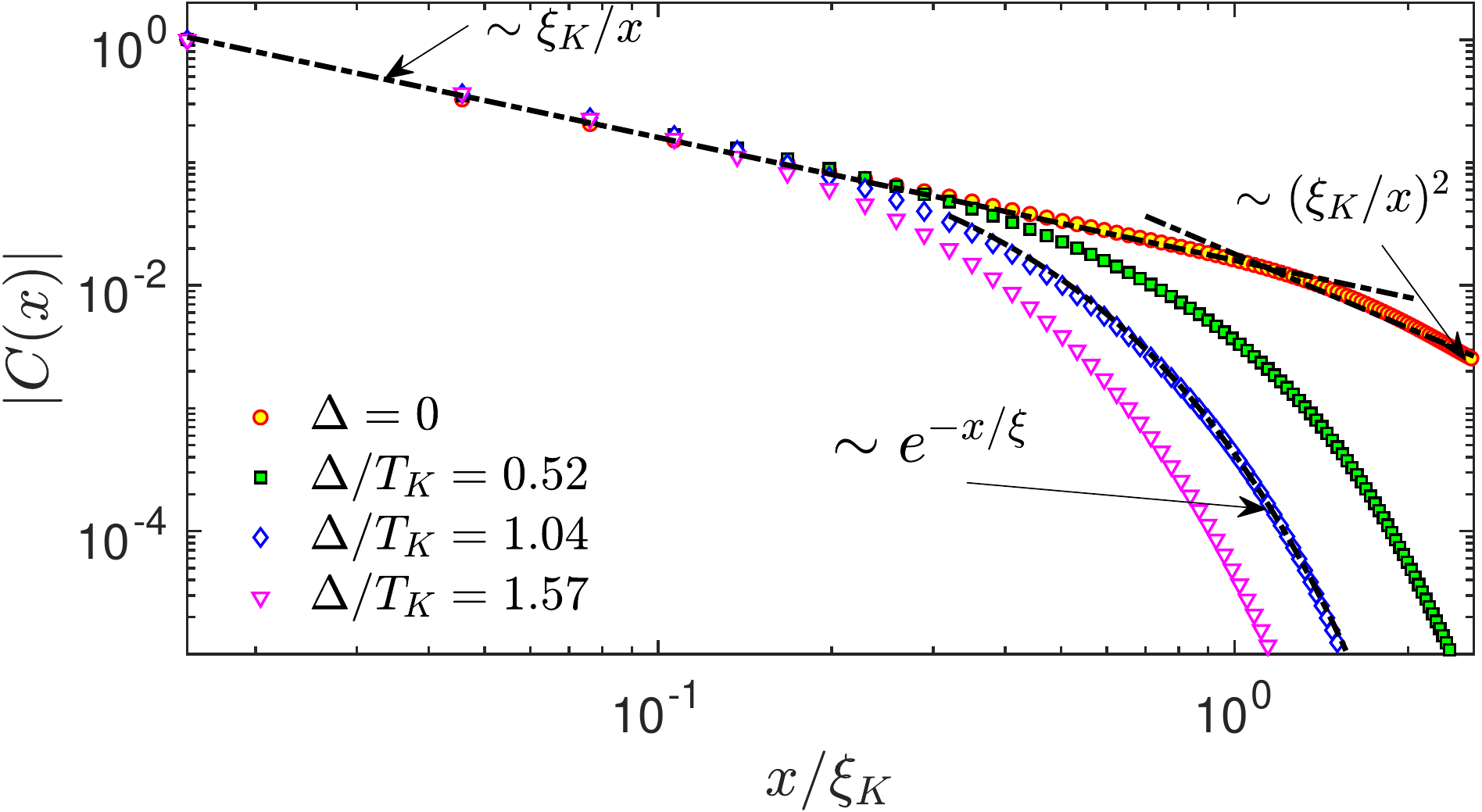}
	\caption{Envelope for the equal-time spin-spin correlation function $C(x)$ defined in Eq.~\eqref{eq:chi} as a function of the distance from the impurity spin in a one dimensional chain, as  computed by DMRG.
	For $\Delta = 0$, the envelope function shows the expected universal scaling 
	in the near and far regions. For $\Delta \ne 0$,  the algebraic behavior  turns into an exponential decay, $C(x)\propto \exp(-x/\xi)$, 
	with $\xi = v_F/\Delta$ the coherence length.}
	\label{fig:correlation}
\end{figure}

This behavior is clearly observed  in our DMRG simulations performed on a one-dimensional superconducting lattice, 
with a Kondo  impurity placed at its end  (see Fig.~\ref{fig:correlation}). In our simulations, we focused on the case of half filling,
and extracted the envelope function of $C(x)$ from the value of $C(x) \equiv \average{\bS_\imp\, \cdot \bs(x)}$
at the even sites~\footnote{The Fermi momentum $k_F=\pi/2$, guarantees that the 
even sites  corresponding to $ x = n \pi/k_F$ represents the envelope function itself}. 
For $\Delta =0$,  the envelope function shows the expected behavior of 
$C(x) \sim 1/x $ and  $C(x) \sim 1/x^2$ for small and large distances, respectively.
The presence of the superconducting gap alters this behavior fundamentally, and induces an 
exponential decay of the form, $C(x)\propto \exp(-x/\xi)$, 
once $x$ gets larger than $\xi$. 

\paragraph{Connection to  experiments --}
\begin{figure}[b!]
	\includegraphics[width=0.5\columnwidth]{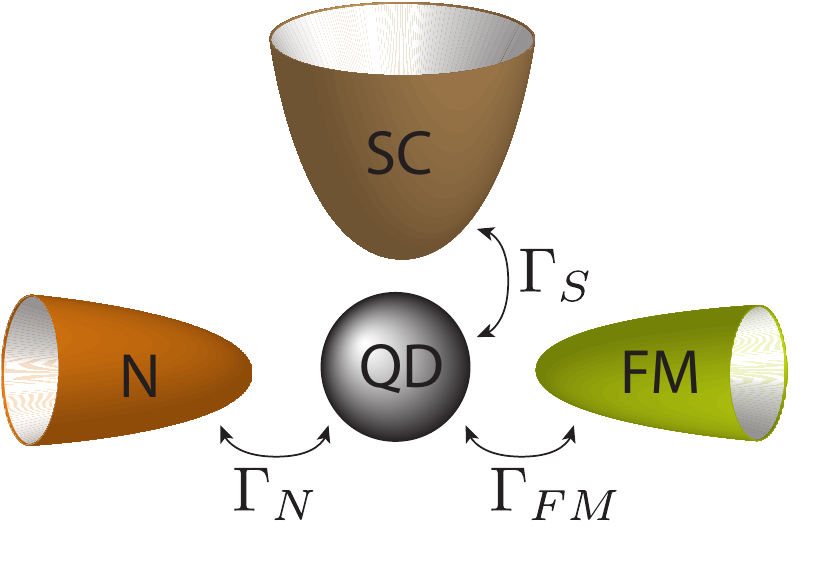}
	\caption{ Experimental setup  for measuring the compensation $\kappa$.
		The quantum dot is coupled to  a normal (N), a superconducting  (SC), and a ferromagnetic (FM) lead. }
	\label{fig:experimental_setup}
\end{figure}   
These predictions  can be tested experimentally. The degree of compensation, in particular, can be measured 
by investigating the magnetic  splitting of an artificial atom (quantum dot), attached to a superconductor, 
and placed in  a local field, as realized in the setup presented in
Fig.~\ref{fig:experimental_setup}. The local exchange field is induced by attaching 
a ferromagnetic electrode to the quantum dot, and the strength of this field can be tuned efficiently 
by shifting the quantum dot's level \cite{Martinek2003,Hauptmann.2008,Hofstetter.2010, Gaass.2011}. A tunnel 
coupling to the superconductor establishes the exchange coupling, $J$, and  gives 
rise to Kondo screening~\cite{Buitelaar.2002,Hofstetter.2010,Deacon.2010, Deacon.2010b}. Finally the third, normal  electrode 
is used  used to perform co-tunneling spectroscopy~\cite{Scherubl.2020}
and thus measure the exchange field induced splitting~\cite{Hauptmann.2008,Hofstetter.2010, Gaass.2011}.
All elements of this circuit have been demonstrated experimentally. 

\paragraph{Conclusions.---} 
We have investigated the fate of the Kondo
cloud of a magnetic impurity embedded in a superconducting host, 
and have shown that the {impurity's} spin remains partially compensated by quantum fluctuations 
even in the superconducting phase. The extension of the fractional compensation cloud is  just the 
superconducting correlation length, $\xi$.   The degree of compensation displays a universal jump at the 
parity changing transition point, and  is a universal function 
of $\Delta/T_K$, which we determined analytically and numerically, and which can be accessed experimentally 
by a proposed experimental set-up. 

\emph{Acknowledgments.--}  
This research has been supported by the National Research Development and Innovation Office (NKFIH) through the OTKA Grant FK 132146, 
 the Hungarian Quantum Technology National Excellence Program under project no. 2017-1.2.1-NKP-2017-00001, and by the NKFIH fund TKP2020 IES (Grant No. BME-IE-NAT), under the auspices of the Ministry for Innovation and Technology.
Support by the Polish National Science Centre grant No. 2017/27/B/ST3/00621,
and by the Romanian National Authority for Scientific Research and Innovation, UEFISCDI, 
under project no. PN-III-P4-ID-PCE-2020-0277 is also acknowledged.

\bibliography{references}

\begin{thebibliography}{59}%
\makeatletter
\providecommand \@ifxundefined [1]{%
 \@ifx{#1\undefined}
}%
\providecommand \@ifnum [1]{%
 \ifnum #1\expandafter \@firstoftwo
 \else \expandafter \@secondoftwo
 \fi
}%
\providecommand \@ifx [1]{%
 \ifx #1\expandafter \@firstoftwo
 \else \expandafter \@secondoftwo
 \fi
}%
\providecommand \natexlab [1]{#1}%
\providecommand \enquote  [1]{``#1''}%
\providecommand \bibnamefont  [1]{#1}%
\providecommand \bibfnamefont [1]{#1}%
\providecommand \citenamefont [1]{#1}%
\providecommand \href@noop [0]{\@secondoftwo}%
\providecommand \href [0]{\begingroup \@sanitize@url \@href}%
\providecommand \@href[1]{\@@startlink{#1}\@@href}%
\providecommand \@@href[1]{\endgroup#1\@@endlink}%
\providecommand \@sanitize@url [0]{\catcode `\\12\catcode `\$12\catcode
  `\&12\catcode `\#12\catcode `\^12\catcode `\_12\catcode `\%12\relax}%
\providecommand \@@startlink[1]{}%
\providecommand \@@endlink[0]{}%
\providecommand \url  [0]{\begingroup\@sanitize@url \@url }%
\providecommand \@url [1]{\endgroup\@href {#1}{\urlprefix }}%
\providecommand \urlprefix  [0]{URL }%
\providecommand \Eprint [0]{\href }%
\providecommand \doibase [0]{http://dx.doi.org/}%
\providecommand \selectlanguage [0]{\@gobble}%
\providecommand \bibinfo  [0]{\@secondoftwo}%
\providecommand \bibfield  [0]{\@secondoftwo}%
\providecommand \translation [1]{[#1]}%
\providecommand \BibitemOpen [0]{}%
\providecommand \bibitemStop [0]{}%
\providecommand \bibitemNoStop [0]{.\EOS\space}%
\providecommand \EOS [0]{\spacefactor3000\relax}%
\providecommand \BibitemShut  [1]{\csname bibitem#1\endcsname}%
\let\auto@bib@innerbib\@empty
\bibitem [{\citenamefont {Hewson}(1992)}]{Hewson.1992}%
  \BibitemOpen
  \bibfield  {author} {\bibinfo {author} {\bibfnamefont {A.~C.}\ \bibnamefont
  {Hewson}},\ }\href@noop {} {\emph {\bibinfo {title} {The Kondo Problem to
  Heavy Fermions}}}\ (\bibinfo  {publisher} {Cambridge University Press},\
  \bibinfo {year} {1992})\BibitemShut {NoStop}%
\bibitem [{\citenamefont {Gubernatis}\ \emph {et~al.}(1987)\citenamefont
  {Gubernatis}, \citenamefont {Hirsch},\ and\ \citenamefont
  {Scalapino}}]{Scalapino.1987}%
  \BibitemOpen
  \bibfield  {author} {\bibinfo {author} {\bibfnamefont {J.~E.}\ \bibnamefont
  {Gubernatis}}, \bibinfo {author} {\bibfnamefont {J.~E.}\ \bibnamefont
  {Hirsch}}, \ and\ \bibinfo {author} {\bibfnamefont {D.~J.}\ \bibnamefont
  {Scalapino}},\ }\href {\doibase 10.1103/PhysRevB.35.8478} {\bibfield
  {journal} {\bibinfo  {journal} {Phys. Rev. B}\ }\textbf {\bibinfo {volume}
  {35}},\ \bibinfo {pages} {8478} (\bibinfo {year} {1987})}\BibitemShut
  {NoStop}%
\bibitem [{\citenamefont {Chen}\ \emph {et~al.}(1987)\citenamefont {Chen},
  \citenamefont {Jayaprakash},\ and\ \citenamefont
  {Krishna-Murthy}}]{Chen.1987}%
  \BibitemOpen
  \bibfield  {author} {\bibinfo {author} {\bibfnamefont {K.}~\bibnamefont
  {Chen}}, \bibinfo {author} {\bibfnamefont {C.}~\bibnamefont {Jayaprakash}}, \
  and\ \bibinfo {author} {\bibfnamefont {H.~R.}\ \bibnamefont
  {Krishna-Murthy}},\ }\href {\doibase 10.1103/PhysRevLett.58.929} {\bibfield
  {journal} {\bibinfo  {journal} {Phys. Rev. Lett.}\ }\textbf {\bibinfo
  {volume} {58}},\ \bibinfo {pages} {929} (\bibinfo {year} {1987})}\BibitemShut
  {NoStop}%
\bibitem [{\citenamefont {Chen}\ \emph {et~al.}(1992)\citenamefont {Chen},
  \citenamefont {Jayaprakash},\ and\ \citenamefont
  {Krishnamurthy}}]{Chen.1992}%
  \BibitemOpen
  \bibfield  {author} {\bibinfo {author} {\bibfnamefont {K.}~\bibnamefont
  {Chen}}, \bibinfo {author} {\bibfnamefont {C.}~\bibnamefont {Jayaprakash}}, \
  and\ \bibinfo {author} {\bibfnamefont {H.~R.}\ \bibnamefont
  {Krishnamurthy}},\ }\href {\doibase 10.1103/PhysRevB.45.5368} {\bibfield
  {journal} {\bibinfo  {journal} {Phys. Rev. B}\ }\textbf {\bibinfo {volume}
  {45}},\ \bibinfo {pages} {5368} (\bibinfo {year} {1992})}\BibitemShut
  {NoStop}%
\bibitem [{\citenamefont {S\o{}rensen}\ and\ \citenamefont
  {Affleck}(1996)}]{Affleck.1996a}%
  \BibitemOpen
  \bibfield  {author} {\bibinfo {author} {\bibfnamefont {E.~S.}\ \bibnamefont
  {S\o{}rensen}}\ and\ \bibinfo {author} {\bibfnamefont {I.}~\bibnamefont
  {Affleck}},\ }\href {\doibase 10.1103/PhysRevB.53.9153} {\bibfield  {journal}
  {\bibinfo  {journal} {Phys. Rev. B}\ }\textbf {\bibinfo {volume} {53}},\
  \bibinfo {pages} {9153} (\bibinfo {year} {1996})}\BibitemShut {NoStop}%
\bibitem [{\citenamefont {Barzykin}\ and\ \citenamefont
  {Affleck}(1996)}]{Affleck.1996b}%
  \BibitemOpen
  \bibfield  {author} {\bibinfo {author} {\bibfnamefont {V.}~\bibnamefont
  {Barzykin}}\ and\ \bibinfo {author} {\bibfnamefont {I.}~\bibnamefont
  {Affleck}},\ }\href {\doibase 10.1103/PhysRevLett.76.4959} {\bibfield
  {journal} {\bibinfo  {journal} {Phys. Rev. Lett.}\ }\textbf {\bibinfo
  {volume} {76}},\ \bibinfo {pages} {4959} (\bibinfo {year}
  {1996})}\BibitemShut {NoStop}%
\bibitem [{\citenamefont {Affleck}\ and\ \citenamefont
  {Simon}(2001)}]{Affleck.2001}%
  \BibitemOpen
  \bibfield  {author} {\bibinfo {author} {\bibfnamefont {I.}~\bibnamefont
  {Affleck}}\ and\ \bibinfo {author} {\bibfnamefont {P.}~\bibnamefont
  {Simon}},\ }\href {\doibase 10.1103/PhysRevLett.86.2854} {\bibfield
  {journal} {\bibinfo  {journal} {Phys. Rev. Lett.}\ }\textbf {\bibinfo
  {volume} {86}},\ \bibinfo {pages} {2854} (\bibinfo {year}
  {2001})}\BibitemShut {NoStop}%
\bibitem [{\citenamefont {Costamagna}\ \emph {et~al.}(2006)\citenamefont
  {Costamagna}, \citenamefont {Gazza}, \citenamefont {Torio},\ and\
  \citenamefont {Riera}}]{Costamagna.2006}%
  \BibitemOpen
  \bibfield  {author} {\bibinfo {author} {\bibfnamefont {S.}~\bibnamefont
  {Costamagna}}, \bibinfo {author} {\bibfnamefont {C.~J.}\ \bibnamefont
  {Gazza}}, \bibinfo {author} {\bibfnamefont {M.~E.}\ \bibnamefont {Torio}}, \
  and\ \bibinfo {author} {\bibfnamefont {J.~A.}\ \bibnamefont {Riera}},\ }\href
  {\doibase 10.1103/PhysRevB.74.195103} {\bibfield  {journal} {\bibinfo
  {journal} {Phys. Rev. B}\ }\textbf {\bibinfo {volume} {74}},\ \bibinfo
  {pages} {195103} (\bibinfo {year} {2006})}\BibitemShut {NoStop}%
\bibitem [{\citenamefont {Hand}\ \emph {et~al.}(2006)\citenamefont {Hand},
  \citenamefont {Kroha},\ and\ \citenamefont {Monien}}]{Hand.2006}%
  \BibitemOpen
  \bibfield  {author} {\bibinfo {author} {\bibfnamefont {T.}~\bibnamefont
  {Hand}}, \bibinfo {author} {\bibfnamefont {J.}~\bibnamefont {Kroha}}, \ and\
  \bibinfo {author} {\bibfnamefont {H.}~\bibnamefont {Monien}},\ }\href
  {\doibase 10.1103/PhysRevLett.97.136604} {\bibfield  {journal} {\bibinfo
  {journal} {Phys. Rev. Lett.}\ }\textbf {\bibinfo {volume} {97}},\ \bibinfo
  {pages} {136604} (\bibinfo {year} {2006})}\BibitemShut {NoStop}%
\bibitem [{\citenamefont {Borda}(2007)}]{Borda.2007}%
  \BibitemOpen
  \bibfield  {author} {\bibinfo {author} {\bibfnamefont {L.}~\bibnamefont
  {Borda}},\ }\href {\doibase 10.1103/PhysRevB.75.041307} {\bibfield  {journal}
  {\bibinfo  {journal} {Phys. Rev. B}\ }\textbf {\bibinfo {volume} {75}},\
  \bibinfo {pages} {041307} (\bibinfo {year} {2007})}\BibitemShut {NoStop}%
\bibitem [{\citenamefont {Affleck}\ \emph {et~al.}(2008)\citenamefont
  {Affleck}, \citenamefont {Borda},\ and\ \citenamefont
  {Saleur}}]{Affleck.2008}%
  \BibitemOpen
  \bibfield  {author} {\bibinfo {author} {\bibfnamefont {I.}~\bibnamefont
  {Affleck}}, \bibinfo {author} {\bibfnamefont {L.}~\bibnamefont {Borda}}, \
  and\ \bibinfo {author} {\bibfnamefont {H.}~\bibnamefont {Saleur}},\ }\href
  {\doibase 10.1103/PhysRevB.77.180404} {\bibfield  {journal} {\bibinfo
  {journal} {Phys. Rev. B}\ }\textbf {\bibinfo {volume} {77}},\ \bibinfo
  {pages} {180404} (\bibinfo {year} {2008})}\BibitemShut {NoStop}%
\bibitem [{\citenamefont {Bergmann}(2008)}]{Bergmann.2008}%
  \BibitemOpen
  \bibfield  {author} {\bibinfo {author} {\bibfnamefont {G.}~\bibnamefont
  {Bergmann}},\ }\href {\doibase 10.1103/PhysRevB.77.104401} {\bibfield
  {journal} {\bibinfo  {journal} {Phys. Rev. B}\ }\textbf {\bibinfo {volume}
  {77}},\ \bibinfo {pages} {104401} (\bibinfo {year} {2008})}\BibitemShut
  {NoStop}%
\bibitem [{\citenamefont {Holzner}\ \emph {et~al.}(2009)\citenamefont
  {Holzner}, \citenamefont {McCulloch}, \citenamefont {Schollw\"ock},
  \citenamefont {von Delft},\ and\ \citenamefont
  {Heidrich-Meisner}}]{Holzner.2009}%
  \BibitemOpen
  \bibfield  {author} {\bibinfo {author} {\bibfnamefont {A.}~\bibnamefont
  {Holzner}}, \bibinfo {author} {\bibfnamefont {I.~P.}\ \bibnamefont
  {McCulloch}}, \bibinfo {author} {\bibfnamefont {U.}~\bibnamefont
  {Schollw\"ock}}, \bibinfo {author} {\bibfnamefont {J.}~\bibnamefont {von
  Delft}}, \ and\ \bibinfo {author} {\bibfnamefont {F.}~\bibnamefont
  {Heidrich-Meisner}},\ }\href {\doibase 10.1103/PhysRevB.80.205114} {\bibfield
   {journal} {\bibinfo  {journal} {Phys. Rev. B}\ }\textbf {\bibinfo {volume}
  {80}},\ \bibinfo {pages} {205114} (\bibinfo {year} {2009})}\BibitemShut
  {NoStop}%
\bibitem [{\citenamefont {B\"usser}\ \emph {et~al.}(2010)\citenamefont
  {B\"usser}, \citenamefont {Martins}, \citenamefont {Costa~Ribeiro},
  \citenamefont {Vernek}, \citenamefont {Anda},\ and\ \citenamefont
  {Dagotto}}]{Dagotto.2010}%
  \BibitemOpen
  \bibfield  {author} {\bibinfo {author} {\bibfnamefont {C.~A.}\ \bibnamefont
  {B\"usser}}, \bibinfo {author} {\bibfnamefont {G.~B.}\ \bibnamefont
  {Martins}}, \bibinfo {author} {\bibfnamefont {L.}~\bibnamefont
  {Costa~Ribeiro}}, \bibinfo {author} {\bibfnamefont {E.}~\bibnamefont
  {Vernek}}, \bibinfo {author} {\bibfnamefont {E.~V.}\ \bibnamefont {Anda}}, \
  and\ \bibinfo {author} {\bibfnamefont {E.}~\bibnamefont {Dagotto}},\ }\href
  {\doibase 10.1103/PhysRevB.81.045111} {\bibfield  {journal} {\bibinfo
  {journal} {Phys. Rev. B}\ }\textbf {\bibinfo {volume} {81}},\ \bibinfo
  {pages} {045111} (\bibinfo {year} {2010})}\BibitemShut {NoStop}%
\bibitem [{\citenamefont {Mitchell}\ \emph {et~al.}(2011)\citenamefont
  {Mitchell}, \citenamefont {Becker},\ and\ \citenamefont
  {Bulla}}]{Mitchell.2011}%
  \BibitemOpen
  \bibfield  {author} {\bibinfo {author} {\bibfnamefont {A.~K.}\ \bibnamefont
  {Mitchell}}, \bibinfo {author} {\bibfnamefont {M.}~\bibnamefont {Becker}}, \
  and\ \bibinfo {author} {\bibfnamefont {R.}~\bibnamefont {Bulla}},\ }\href
  {\doibase 10.1103/PhysRevB.84.115120} {\bibfield  {journal} {\bibinfo
  {journal} {Phys. Rev. B}\ }\textbf {\bibinfo {volume} {84}},\ \bibinfo
  {pages} {115120} (\bibinfo {year} {2011})}\BibitemShut {NoStop}%
\bibitem [{\citenamefont {Medvedyeva}\ \emph {et~al.}(2013)\citenamefont
  {Medvedyeva}, \citenamefont {Hoffmann},\ and\ \citenamefont
  {Kehrein}}]{Medvedyeva.2013}%
  \BibitemOpen
  \bibfield  {author} {\bibinfo {author} {\bibfnamefont {M.}~\bibnamefont
  {Medvedyeva}}, \bibinfo {author} {\bibfnamefont {A.}~\bibnamefont
  {Hoffmann}}, \ and\ \bibinfo {author} {\bibfnamefont {S.}~\bibnamefont
  {Kehrein}},\ }\href {\doibase 10.1103/PhysRevB.88.094306} {\bibfield
  {journal} {\bibinfo  {journal} {Phys. Rev. B}\ }\textbf {\bibinfo {volume}
  {88}},\ \bibinfo {pages} {094306} (\bibinfo {year} {2013})}\BibitemShut
  {NoStop}%
\bibitem [{\citenamefont {Lechtenberg}\ and\ \citenamefont
  {Anders}(2014)}]{Lechtenberg.2014}%
  \BibitemOpen
  \bibfield  {author} {\bibinfo {author} {\bibfnamefont {B.}~\bibnamefont
  {Lechtenberg}}\ and\ \bibinfo {author} {\bibfnamefont {F.~B.}\ \bibnamefont
  {Anders}},\ }\href {\doibase 10.1103/PhysRevB.90.045117} {\bibfield
  {journal} {\bibinfo  {journal} {Phys. Rev. B}\ }\textbf {\bibinfo {volume}
  {90}},\ \bibinfo {pages} {045117} (\bibinfo {year} {2014})}\BibitemShut
  {NoStop}%
\bibitem [{\citenamefont {Ghosh}\ \emph {et~al.}(2014)\citenamefont {Ghosh},
  \citenamefont {Ribeiro},\ and\ \citenamefont {Haque}}]{Ghosh.2014}%
  \BibitemOpen
  \bibfield  {author} {\bibinfo {author} {\bibfnamefont {S.}~\bibnamefont
  {Ghosh}}, \bibinfo {author} {\bibfnamefont {P.}~\bibnamefont {Ribeiro}}, \
  and\ \bibinfo {author} {\bibfnamefont {M.}~\bibnamefont {Haque}},\ }\href
  {\doibase 10.1088/1742-5468/2014/04/p04011} {\bibfield  {journal} {\bibinfo
  {journal} {Journal of Statistical Mechanics: Theory and Experiment}\ }\textbf
  {\bibinfo {volume} {2014}},\ \bibinfo {pages} {P04011} (\bibinfo {year}
  {2014})}\BibitemShut {NoStop}%
\bibitem [{\citenamefont {Nuss}\ \emph {et~al.}(2015)\citenamefont {Nuss},
  \citenamefont {Ganahl}, \citenamefont {Arrigoni}, \citenamefont {von~der
  Linden},\ and\ \citenamefont {Evertz}}]{Arrigoni.2015}%
  \BibitemOpen
  \bibfield  {author} {\bibinfo {author} {\bibfnamefont {M.}~\bibnamefont
  {Nuss}}, \bibinfo {author} {\bibfnamefont {M.}~\bibnamefont {Ganahl}},
  \bibinfo {author} {\bibfnamefont {E.}~\bibnamefont {Arrigoni}}, \bibinfo
  {author} {\bibfnamefont {W.}~\bibnamefont {von~der Linden}}, \ and\ \bibinfo
  {author} {\bibfnamefont {H.~G.}\ \bibnamefont {Evertz}},\ }\href {\doibase
  10.1103/PhysRevB.91.085127} {\bibfield  {journal} {\bibinfo  {journal} {Phys.
  Rev. B}\ }\textbf {\bibinfo {volume} {91}},\ \bibinfo {pages} {085127}
  (\bibinfo {year} {2015})}\BibitemShut {NoStop}%
\bibitem [{\citenamefont {Florens}\ and\ \citenamefont
  {Snyman}(2015)}]{Florens.2015}%
  \BibitemOpen
  \bibfield  {author} {\bibinfo {author} {\bibfnamefont {S.}~\bibnamefont
  {Florens}}\ and\ \bibinfo {author} {\bibfnamefont {I.}~\bibnamefont
  {Snyman}},\ }\href {\doibase 10.1103/PhysRevB.92.195106} {\bibfield
  {journal} {\bibinfo  {journal} {Phys. Rev. B}\ }\textbf {\bibinfo {volume}
  {92}},\ \bibinfo {pages} {195106} (\bibinfo {year} {2015})}\BibitemShut
  {NoStop}%
\bibitem [{\citenamefont {Boyce}\ and\ \citenamefont
  {Slichter}(1974)}]{Boyce.1974}%
  \BibitemOpen
  \bibfield  {author} {\bibinfo {author} {\bibfnamefont {J.~B.}\ \bibnamefont
  {Boyce}}\ and\ \bibinfo {author} {\bibfnamefont {C.~P.}\ \bibnamefont
  {Slichter}},\ }\href {\doibase 10.1103/PhysRevLett.32.61} {\bibfield
  {journal} {\bibinfo  {journal} {Phys. Rev. Lett.}\ }\textbf {\bibinfo
  {volume} {32}},\ \bibinfo {pages} {61} (\bibinfo {year} {1974})}\BibitemShut
  {NoStop}%
\bibitem [{\citenamefont {Pr{\"u}ser}\ \emph {et~al.}(2011)\citenamefont
  {Pr{\"u}ser}, \citenamefont {Wenderoth}, \citenamefont {Dargel},
  \citenamefont {Weismann}, \citenamefont {Peters}, \citenamefont {Pruschke},\
  and\ \citenamefont {Ulbrich}}]{Pruser.2011}%
  \BibitemOpen
  \bibfield  {author} {\bibinfo {author} {\bibfnamefont {H.}~\bibnamefont
  {Pr{\"u}ser}}, \bibinfo {author} {\bibfnamefont {M.}~\bibnamefont
  {Wenderoth}}, \bibinfo {author} {\bibfnamefont {P.~E.}\ \bibnamefont
  {Dargel}}, \bibinfo {author} {\bibfnamefont {A.}~\bibnamefont {Weismann}},
  \bibinfo {author} {\bibfnamefont {R.}~\bibnamefont {Peters}}, \bibinfo
  {author} {\bibfnamefont {T.}~\bibnamefont {Pruschke}}, \ and\ \bibinfo
  {author} {\bibfnamefont {R.~G.}\ \bibnamefont {Ulbrich}},\ }\href {\doibase
  10.1038/nphys1876} {\bibfield  {journal} {\bibinfo  {journal} {Nature
  Physics}\ }\textbf {\bibinfo {volume} {7}},\ \bibinfo {pages} {203} (\bibinfo
  {year} {2011})}\BibitemShut {NoStop}%
\bibitem [{\citenamefont {Jiang}\ \emph {et~al.}(2011)\citenamefont {Jiang},
  \citenamefont {Zhang}, \citenamefont {Cao}, \citenamefont {Wu},\ and\
  \citenamefont {Ho}}]{Jiang.2011}%
  \BibitemOpen
  \bibfield  {author} {\bibinfo {author} {\bibfnamefont {Y.}~\bibnamefont
  {Jiang}}, \bibinfo {author} {\bibfnamefont {Y.~N.}\ \bibnamefont {Zhang}},
  \bibinfo {author} {\bibfnamefont {J.~X.}\ \bibnamefont {Cao}}, \bibinfo
  {author} {\bibfnamefont {R.~Q.}\ \bibnamefont {Wu}}, \ and\ \bibinfo {author}
  {\bibfnamefont {W.}~\bibnamefont {Ho}},\ }\href {\doibase
  10.1126/science.1205785} {\bibfield  {journal} {\bibinfo  {journal}
  {Science}\ }\textbf {\bibinfo {volume} {333}},\ \bibinfo {pages} {324}
  (\bibinfo {year} {2011})}\BibitemShut {NoStop}%
\bibitem [{\citenamefont {Figgins}\ \emph {et~al.}(2019)\citenamefont
  {Figgins}, \citenamefont {Mattos}, \citenamefont {Mar}, \citenamefont {Chen},
  \citenamefont {Manoharan},\ and\ \citenamefont {Morr}}]{Figgins.2019}%
  \BibitemOpen
  \bibfield  {author} {\bibinfo {author} {\bibfnamefont {J.}~\bibnamefont
  {Figgins}}, \bibinfo {author} {\bibfnamefont {L.~S.}\ \bibnamefont {Mattos}},
  \bibinfo {author} {\bibfnamefont {W.}~\bibnamefont {Mar}}, \bibinfo {author}
  {\bibfnamefont {Y.-T.}\ \bibnamefont {Chen}}, \bibinfo {author}
  {\bibfnamefont {H.~C.}\ \bibnamefont {Manoharan}}, \ and\ \bibinfo {author}
  {\bibfnamefont {D.~K.}\ \bibnamefont {Morr}},\ }\href {\doibase
  10.1038/s41467-019-13446-1} {\bibfield  {journal} {\bibinfo  {journal}
  {Nature Communications}\ }\textbf {\bibinfo {volume} {10}},\ \bibinfo {pages}
  {5588} (\bibinfo {year} {2019})}\BibitemShut {NoStop}%
\bibitem [{Note1()}]{Note1}%
  \BibitemOpen
  \bibinfo {note} {We use units $\hbar = k_B = 1$.}\BibitemShut {Stop}%
\bibitem [{\citenamefont {Affleck}(2010)}]{Affleck.2010}%
  \BibitemOpen
  \bibfield  {author} {\bibinfo {author} {\bibfnamefont {I.}~\bibnamefont
  {Affleck}},\ }\href@noop {} {\enquote {\bibinfo {title} {The kondo screening
  cloud: what it is and how to observe it},}\ } (\bibinfo {year} {2010}),\
  \Eprint {http://arxiv.org/abs/0911.2209} {arXiv:0911.2209
  [cond-mat.mes-hall]} \BibitemShut {NoStop}%
\bibitem [{\citenamefont {Borda}\ \emph {et~al.}(2009)\citenamefont {Borda},
  \citenamefont {Garst},\ and\ \citenamefont {Kroha}}]{Borda.2009}%
  \BibitemOpen
  \bibfield  {author} {\bibinfo {author} {\bibfnamefont {L.}~\bibnamefont
  {Borda}}, \bibinfo {author} {\bibfnamefont {M.}~\bibnamefont {Garst}}, \ and\
  \bibinfo {author} {\bibfnamefont {J.}~\bibnamefont {Kroha}},\ }\href
  {\doibase 10.1103/PhysRevB.79.100408} {\bibfield  {journal} {\bibinfo
  {journal} {Phys. Rev. B}\ }\textbf {\bibinfo {volume} {79}},\ \bibinfo
  {pages} {100408} (\bibinfo {year} {2009})}\BibitemShut {NoStop}%
\bibitem [{\citenamefont {Bayat}\ \emph {et~al.}(2010)\citenamefont {Bayat},
  \citenamefont {Sodano},\ and\ \citenamefont {Bose}}]{Bayat.2010}%
  \BibitemOpen
  \bibfield  {author} {\bibinfo {author} {\bibfnamefont {A.}~\bibnamefont
  {Bayat}}, \bibinfo {author} {\bibfnamefont {P.}~\bibnamefont {Sodano}}, \
  and\ \bibinfo {author} {\bibfnamefont {S.}~\bibnamefont {Bose}},\ }\href
  {\doibase 10.1103/PhysRevB.81.064429} {\bibfield  {journal} {\bibinfo
  {journal} {Phys. Rev. B}\ }\textbf {\bibinfo {volume} {81}},\ \bibinfo
  {pages} {064429} (\bibinfo {year} {2010})}\BibitemShut {NoStop}%
\bibitem [{\citenamefont {Bayat}\ \emph {et~al.}(2012)\citenamefont {Bayat},
  \citenamefont {Bose}, \citenamefont {Sodano},\ and\ \citenamefont
  {Johannesson}}]{Bayat.2012}%
  \BibitemOpen
  \bibfield  {author} {\bibinfo {author} {\bibfnamefont {A.}~\bibnamefont
  {Bayat}}, \bibinfo {author} {\bibfnamefont {S.}~\bibnamefont {Bose}},
  \bibinfo {author} {\bibfnamefont {P.}~\bibnamefont {Sodano}}, \ and\ \bibinfo
  {author} {\bibfnamefont {H.}~\bibnamefont {Johannesson}},\ }\href {\doibase
  10.1103/PhysRevLett.109.066403} {\bibfield  {journal} {\bibinfo  {journal}
  {Phys. Rev. Lett.}\ }\textbf {\bibinfo {volume} {109}},\ \bibinfo {pages}
  {066403} (\bibinfo {year} {2012})}\BibitemShut {NoStop}%
\bibitem [{\citenamefont {Lee}\ \emph {et~al.}(2015)\citenamefont {Lee},
  \citenamefont {Park},\ and\ \citenamefont {Sim}}]{Lee.2015}%
  \BibitemOpen
  \bibfield  {author} {\bibinfo {author} {\bibfnamefont {S.-S.~B.}\
  \bibnamefont {Lee}}, \bibinfo {author} {\bibfnamefont {J.}~\bibnamefont
  {Park}}, \ and\ \bibinfo {author} {\bibfnamefont {H.-S.}\ \bibnamefont
  {Sim}},\ }\href {\doibase 10.1103/PhysRevLett.114.057203} {\bibfield
  {journal} {\bibinfo  {journal} {Phys. Rev. Lett.}\ }\textbf {\bibinfo
  {volume} {114}},\ \bibinfo {pages} {057203} (\bibinfo {year}
  {2015})}\BibitemShut {NoStop}%
\bibitem [{\citenamefont {Park}\ \emph {et~al.}(2013)\citenamefont {Park},
  \citenamefont {Lee}, \citenamefont {Oreg},\ and\ \citenamefont
  {Sim}}]{Yuval.2013}%
  \BibitemOpen
  \bibfield  {author} {\bibinfo {author} {\bibfnamefont {J.}~\bibnamefont
  {Park}}, \bibinfo {author} {\bibfnamefont {S.-S.~B.}\ \bibnamefont {Lee}},
  \bibinfo {author} {\bibfnamefont {Y.}~\bibnamefont {Oreg}}, \ and\ \bibinfo
  {author} {\bibfnamefont {H.-S.}\ \bibnamefont {Sim}},\ }\href {\doibase
  10.1103/PhysRevLett.110.246603} {\bibfield  {journal} {\bibinfo  {journal}
  {Phys. Rev. Lett.}\ }\textbf {\bibinfo {volume} {110}},\ \bibinfo {pages}
  {246603} (\bibinfo {year} {2013})}\BibitemShut {NoStop}%
\bibitem [{\citenamefont {V.~Borzenets}\ \emph {et~al.}(2020)\citenamefont
  {V.~Borzenets}, \citenamefont {Shim}, \citenamefont {Chen}, \citenamefont
  {Ludwig}, \citenamefont {Wieck}, \citenamefont {Tarucha}, \citenamefont
  {Sim},\ and\ \citenamefont {Yamamoto}}]{Tarucha.2020}%
  \BibitemOpen
  \bibfield  {author} {\bibinfo {author} {\bibfnamefont {I.}~\bibnamefont
  {V.~Borzenets}}, \bibinfo {author} {\bibfnamefont {J.}~\bibnamefont {Shim}},
  \bibinfo {author} {\bibfnamefont {J.~C.~H.}\ \bibnamefont {Chen}}, \bibinfo
  {author} {\bibfnamefont {A.}~\bibnamefont {Ludwig}}, \bibinfo {author}
  {\bibfnamefont {A.~D.}\ \bibnamefont {Wieck}}, \bibinfo {author}
  {\bibfnamefont {S.}~\bibnamefont {Tarucha}}, \bibinfo {author} {\bibfnamefont
  {H.-S.}\ \bibnamefont {Sim}}, \ and\ \bibinfo {author} {\bibfnamefont
  {M.}~\bibnamefont {Yamamoto}},\ }\href
  {https://doi.org/10.1038/s41586-020-2058-6} {\bibfield  {journal} {\bibinfo
  {journal} {Nature}\ }\textbf {\bibinfo {volume} {579}},\ \bibinfo {pages}
  {210} (\bibinfo {year} {2020})}\BibitemShut {NoStop}%
\bibitem [{Note2()}]{Note2}%
  \BibitemOpen
  \bibinfo {note} {These two Shiba states are not distinct, they are the
  electronic and hole parts of the same quantum state}\BibitemShut {NoStop}%
\bibitem [{\citenamefont {Yu}(1965)}]{Yu.1965}%
  \BibitemOpen
  \bibfield  {author} {\bibinfo {author} {\bibfnamefont {L.}~\bibnamefont
  {Yu}},\ }\href {http://wulixb.iphy.ac.cn/en/article/id/851} {\bibfield
  {journal} {\bibinfo  {journal} {Acta Physica Sinica}\ }\textbf {\bibinfo
  {volume} {21}},\ \bibinfo {pages} {115304} (\bibinfo {year}
  {1965})}\BibitemShut {NoStop}%
\bibitem [{\citenamefont {Shiba}(1968)}]{Shiba.1968}%
  \BibitemOpen
  \bibfield  {author} {\bibinfo {author} {\bibfnamefont {H.}~\bibnamefont
  {Shiba}},\ }\href {https://academic.oup.com/ptp/article/40/3/435/1831894#}
  {\bibfield  {journal} {\bibinfo  {journal} {Progress of theoretical Physics}\
  }\textbf {\bibinfo {volume} {40}},\ \bibinfo {pages} {435} (\bibinfo {year}
  {1968})}\BibitemShut {NoStop}%
\bibitem [{\citenamefont {Rusinov}\ \emph {et~al.}(1974)\citenamefont
  {Rusinov}, \citenamefont {Kat},\ and\ \citenamefont {Kopaev}}]{Rusinov.1974}%
  \BibitemOpen
  \bibfield  {author} {\bibinfo {author} {\bibfnamefont {A.~I.}\ \bibnamefont
  {Rusinov}}, \bibinfo {author} {\bibfnamefont {D.~C.}\ \bibnamefont {Kat}}, \
  and\ \bibinfo {author} {\bibfnamefont {Y.~V.}\ \bibnamefont {Kopaev}},\
  }\href {http://www.jetp.ac.ru/cgi-bin/e/index/e/38/5/p991?a=lists} {\bibfield
   {journal} {\bibinfo  {journal} {Journal of Experimental and Theoretical
  Physics}\ }\textbf {\bibinfo {volume} {38}},\ \bibinfo {pages} {991}
  (\bibinfo {year} {1974})}\BibitemShut {NoStop}%
\bibitem [{Note3()}]{Note3}%
  \BibitemOpen
  \bibinfo {note} {The critical value depends on the precise definition of the
  Kondo temperature, $T_K$. Throughout this work, we define $T_K$ as the
  half-width of the Kondo resonance, i.e., that of the composite fermion's
  spectral function~\cite {Costi.2000}.}\BibitemShut {Stop}%
\bibitem [{\citenamefont {Scher{\"u}bl}\ \emph {et~al.}(2020)\citenamefont
  {Scher{\"u}bl}, \citenamefont {F{\"u}l{\"o}p}, \citenamefont {Moca},
  \citenamefont {Gramich}, \citenamefont {Baumgartner}, \citenamefont {Makk},
  \citenamefont {Elalaily}, \citenamefont {Sch{\"o}nenberger}, \citenamefont
  {Nyg{\aa}rd}, \citenamefont {Zar{\'a}nd},\ and\ \citenamefont
  {Csonka}}]{Scherubl.2020}%
  \BibitemOpen
  \bibfield  {author} {\bibinfo {author} {\bibfnamefont {Z.}~\bibnamefont
  {Scher{\"u}bl}}, \bibinfo {author} {\bibfnamefont {G.}~\bibnamefont
  {F{\"u}l{\"o}p}}, \bibinfo {author} {\bibfnamefont {C.~P.}\ \bibnamefont
  {Moca}}, \bibinfo {author} {\bibfnamefont {J.}~\bibnamefont {Gramich}},
  \bibinfo {author} {\bibfnamefont {A.}~\bibnamefont {Baumgartner}}, \bibinfo
  {author} {\bibfnamefont {P.}~\bibnamefont {Makk}}, \bibinfo {author}
  {\bibfnamefont {T.}~\bibnamefont {Elalaily}}, \bibinfo {author}
  {\bibfnamefont {C.}~\bibnamefont {Sch{\"o}nenberger}}, \bibinfo {author}
  {\bibfnamefont {J.}~\bibnamefont {Nyg{\aa}rd}}, \bibinfo {author}
  {\bibfnamefont {G.}~\bibnamefont {Zar{\'a}nd}}, \ and\ \bibinfo {author}
  {\bibfnamefont {S.}~\bibnamefont {Csonka}},\ }\href {\doibase
  10.1038/s41467-020-15322-9} {\bibfield  {journal} {\bibinfo  {journal}
  {Nature Communications}\ }\textbf {\bibinfo {volume} {11}},\ \bibinfo {pages}
  {1834} (\bibinfo {year} {2020})}\BibitemShut {NoStop}%
\bibitem [{\citenamefont {Moca}\ \emph {et~al.}()\citenamefont {Moca},
  \citenamefont {Weymann}, \citenamefont {Werner},\ and\ \citenamefont
  {Zarand}}]{SM}%
  \BibitemOpen
  \bibfield  {author} {\bibinfo {author} {\bibfnamefont {C.~P.}\ \bibnamefont
  {Moca}}, \bibinfo {author} {\bibfnamefont {I.}~\bibnamefont {Weymann}},
  \bibinfo {author} {\bibfnamefont {M.~A.}\ \bibnamefont {Werner}}, \ and\
  \bibinfo {author} {\bibfnamefont {G.}~\bibnamefont {Zarand}},\ }\href@noop {}
  {\bibinfo  {journal} {Supplementary Material}\ }\BibitemShut {NoStop}%
\bibitem [{\citenamefont {Costi}(2000)}]{Costi.2000}%
  \BibitemOpen
\bibfield  {journal} {  }\bibfield  {author} {\bibinfo {author} {\bibfnamefont
  {T.~A.}\ \bibnamefont {Costi}},\ }\href {\doibase
  10.1103/PhysRevLett.85.1504} {\bibfield  {journal} {\bibinfo  {journal}
  {Phys. Rev. Lett.}\ }\textbf {\bibinfo {volume} {85}},\ \bibinfo {pages}
  {1504} (\bibinfo {year} {2000})}\BibitemShut {NoStop}%
\bibitem [{\citenamefont {Wilson}(1975)}]{Wilson.1975}%
  \BibitemOpen
  \bibfield  {author} {\bibinfo {author} {\bibfnamefont {K.~G.}\ \bibnamefont
  {Wilson}},\ }\href {\doibase 10.1103/RevModPhys.47.773} {\bibfield  {journal}
  {\bibinfo  {journal} {Rev. Mod. Phys.}\ }\textbf {\bibinfo {volume} {47}},\
  \bibinfo {pages} {773} (\bibinfo {year} {1975})}\BibitemShut {NoStop}%
\bibitem [{\citenamefont {White}(1996)}]{White.1996}%
  \BibitemOpen
  \bibfield  {author} {\bibinfo {author} {\bibfnamefont {S.~R.}\ \bibnamefont
  {White}},\ }\href {\doibase 10.1103/PhysRevLett.77.3633} {\bibfield
  {journal} {\bibinfo  {journal} {Phys. Rev. Lett.}\ }\textbf {\bibinfo
  {volume} {77}},\ \bibinfo {pages} {3633} (\bibinfo {year}
  {1996})}\BibitemShut {NoStop}%
\bibitem [{\citenamefont {Schollw\"ock}(2005)}]{Schollwock.2005}%
  \BibitemOpen
  \bibfield  {author} {\bibinfo {author} {\bibfnamefont {U.}~\bibnamefont
  {Schollw\"ock}},\ }\href {\doibase 10.1103/RevModPhys.77.259} {\bibfield
  {journal} {\bibinfo  {journal} {Rev. Mod. Phys.}\ }\textbf {\bibinfo {volume}
  {77}},\ \bibinfo {pages} {259} (\bibinfo {year} {2005})}\BibitemShut
  {NoStop}%
\bibitem [{\citenamefont {Ishii}(1978)}]{Ishii.1978}%
  \BibitemOpen
  \bibfield  {author} {\bibinfo {author} {\bibfnamefont {H.}~\bibnamefont
  {Ishii}},\ }\href {\doibase 10.1007/BF00117963} {\bibfield  {journal}
  {\bibinfo  {journal} {Journal of Low Temperature Physics}\ }\textbf {\bibinfo
  {volume} {32}},\ \bibinfo {pages} {457} (\bibinfo {year} {1978})}\BibitemShut
  {NoStop}%
\bibitem [{Note4()}]{Note4}%
  \BibitemOpen
  \bibinfo {note} {The Fermi momentum $k_F=\pi /2$, guarantees that the even
  sites corresponding to $ x = n \pi /k_F$ represents the envelope function
  itself}\BibitemShut {NoStop}%
\bibitem [{\citenamefont {Martinek}\ \emph {et~al.}(2003)\citenamefont
  {Martinek}, \citenamefont {Utsumi}, \citenamefont {Imamura}, \citenamefont
  {Barna\ifmmode~\acute{s}\else \'{s}\fi{}}, \citenamefont {Maekawa},
  \citenamefont {K\"onig},\ and\ \citenamefont {Sch\"on}}]{Martinek2003}%
  \BibitemOpen
  \bibfield  {author} {\bibinfo {author} {\bibfnamefont {J.}~\bibnamefont
  {Martinek}}, \bibinfo {author} {\bibfnamefont {Y.}~\bibnamefont {Utsumi}},
  \bibinfo {author} {\bibfnamefont {H.}~\bibnamefont {Imamura}}, \bibinfo
  {author} {\bibfnamefont {J.}~\bibnamefont {Barna\ifmmode~\acute{s}\else
  \'{s}\fi{}}}, \bibinfo {author} {\bibfnamefont {S.}~\bibnamefont {Maekawa}},
  \bibinfo {author} {\bibfnamefont {J.}~\bibnamefont {K\"onig}}, \ and\
  \bibinfo {author} {\bibfnamefont {G.}~\bibnamefont {Sch\"on}},\ }\href
  {\doibase 10.1103/PhysRevLett.91.127203} {\bibfield  {journal} {\bibinfo
  {journal} {Phys. Rev. Lett.}\ }\textbf {\bibinfo {volume} {91}},\ \bibinfo
  {pages} {127203} (\bibinfo {year} {2003})}\BibitemShut {NoStop}%
\bibitem [{\citenamefont {Hauptmann}\ \emph {et~al.}(2008)\citenamefont
  {Hauptmann}, \citenamefont {Paaske},\ and\ \citenamefont
  {Lindelof}}]{Hauptmann.2008}%
  \BibitemOpen
  \bibfield  {author} {\bibinfo {author} {\bibfnamefont {J.~R.}\ \bibnamefont
  {Hauptmann}}, \bibinfo {author} {\bibfnamefont {J.}~\bibnamefont {Paaske}}, \
  and\ \bibinfo {author} {\bibfnamefont {P.~E.}\ \bibnamefont {Lindelof}},\
  }\href {\doibase 10.1038/nphys931} {\bibfield  {journal} {\bibinfo  {journal}
  {Nature Physics}\ }\textbf {\bibinfo {volume} {4}},\ \bibinfo {pages} {373}
  (\bibinfo {year} {2008})}\BibitemShut {NoStop}%
\bibitem [{\citenamefont {Hofstetter}\ \emph {et~al.}(2010)\citenamefont
  {Hofstetter}, \citenamefont {Geresdi}, \citenamefont {Aagesen}, \citenamefont
  {Nyg\aa{}rd}, \citenamefont {Sch\"onenberger},\ and\ \citenamefont
  {Csonka}}]{Hofstetter.2010}%
  \BibitemOpen
  \bibfield  {author} {\bibinfo {author} {\bibfnamefont {L.}~\bibnamefont
  {Hofstetter}}, \bibinfo {author} {\bibfnamefont {A.}~\bibnamefont {Geresdi}},
  \bibinfo {author} {\bibfnamefont {M.}~\bibnamefont {Aagesen}}, \bibinfo
  {author} {\bibfnamefont {J.}~\bibnamefont {Nyg\aa{}rd}}, \bibinfo {author}
  {\bibfnamefont {C.}~\bibnamefont {Sch\"onenberger}}, \ and\ \bibinfo {author}
  {\bibfnamefont {S.}~\bibnamefont {Csonka}},\ }\href {\doibase
  10.1103/PhysRevLett.104.246804} {\bibfield  {journal} {\bibinfo  {journal}
  {Phys. Rev. Lett.}\ }\textbf {\bibinfo {volume} {104}},\ \bibinfo {pages}
  {246804} (\bibinfo {year} {2010})}\BibitemShut {NoStop}%
\bibitem [{\citenamefont {Gaass}\ \emph {et~al.}(2011)\citenamefont {Gaass},
  \citenamefont {H\"uttel}, \citenamefont {Kang}, \citenamefont {Weymann},
  \citenamefont {von Delft},\ and\ \citenamefont {Strunk}}]{Gaass.2011}%
  \BibitemOpen
  \bibfield  {author} {\bibinfo {author} {\bibfnamefont {M.}~\bibnamefont
  {Gaass}}, \bibinfo {author} {\bibfnamefont {A.~K.}\ \bibnamefont {H\"uttel}},
  \bibinfo {author} {\bibfnamefont {K.}~\bibnamefont {Kang}}, \bibinfo {author}
  {\bibfnamefont {I.}~\bibnamefont {Weymann}}, \bibinfo {author} {\bibfnamefont
  {J.}~\bibnamefont {von Delft}}, \ and\ \bibinfo {author} {\bibfnamefont
  {C.}~\bibnamefont {Strunk}},\ }\href {\doibase
  10.1103/PhysRevLett.107.176808} {\bibfield  {journal} {\bibinfo  {journal}
  {Phys. Rev. Lett.}\ }\textbf {\bibinfo {volume} {107}},\ \bibinfo {pages}
  {176808} (\bibinfo {year} {2011})}\BibitemShut {NoStop}%
\bibitem [{\citenamefont {Buitelaar}\ \emph {et~al.}(2002)\citenamefont
  {Buitelaar}, \citenamefont {Nussbaumer},\ and\ \citenamefont
  {Sch\"onenberger}}]{Buitelaar.2002}%
  \BibitemOpen
  \bibfield  {author} {\bibinfo {author} {\bibfnamefont {M.~R.}\ \bibnamefont
  {Buitelaar}}, \bibinfo {author} {\bibfnamefont {T.}~\bibnamefont
  {Nussbaumer}}, \ and\ \bibinfo {author} {\bibfnamefont {C.}~\bibnamefont
  {Sch\"onenberger}},\ }\href {\doibase 10.1103/PhysRevLett.89.256801}
  {\bibfield  {journal} {\bibinfo  {journal} {Phys. Rev. Lett.}\ }\textbf
  {\bibinfo {volume} {89}},\ \bibinfo {pages} {256801} (\bibinfo {year}
  {2002})}\BibitemShut {NoStop}%
\bibitem [{\citenamefont {Deacon}\ \emph
  {et~al.}(2010{\natexlab{a}})\citenamefont {Deacon}, \citenamefont {Tanaka},
  \citenamefont {Oiwa}, \citenamefont {Sakano}, \citenamefont {Yoshida},
  \citenamefont {Shibata}, \citenamefont {Hirakawa},\ and\ \citenamefont
  {Tarucha}}]{Deacon.2010}%
  \BibitemOpen
  \bibfield  {author} {\bibinfo {author} {\bibfnamefont {R.~S.}\ \bibnamefont
  {Deacon}}, \bibinfo {author} {\bibfnamefont {Y.}~\bibnamefont {Tanaka}},
  \bibinfo {author} {\bibfnamefont {A.}~\bibnamefont {Oiwa}}, \bibinfo {author}
  {\bibfnamefont {R.}~\bibnamefont {Sakano}}, \bibinfo {author} {\bibfnamefont
  {K.}~\bibnamefont {Yoshida}}, \bibinfo {author} {\bibfnamefont
  {K.}~\bibnamefont {Shibata}}, \bibinfo {author} {\bibfnamefont
  {K.}~\bibnamefont {Hirakawa}}, \ and\ \bibinfo {author} {\bibfnamefont
  {S.}~\bibnamefont {Tarucha}},\ }\href {\doibase
  10.1103/PhysRevLett.104.076805} {\bibfield  {journal} {\bibinfo  {journal}
  {Phys. Rev. Lett.}\ }\textbf {\bibinfo {volume} {104}},\ \bibinfo {pages}
  {076805} (\bibinfo {year} {2010}{\natexlab{a}})}\BibitemShut {NoStop}%
\bibitem [{\citenamefont {Deacon}\ \emph
  {et~al.}(2010{\natexlab{b}})\citenamefont {Deacon}, \citenamefont {Tanaka},
  \citenamefont {Oiwa}, \citenamefont {Sakano}, \citenamefont {Yoshida},
  \citenamefont {Shibata}, \citenamefont {Hirakawa},\ and\ \citenamefont
  {Tarucha}}]{Deacon.2010b}%
  \BibitemOpen
  \bibfield  {author} {\bibinfo {author} {\bibfnamefont {R.~S.}\ \bibnamefont
  {Deacon}}, \bibinfo {author} {\bibfnamefont {Y.}~\bibnamefont {Tanaka}},
  \bibinfo {author} {\bibfnamefont {A.}~\bibnamefont {Oiwa}}, \bibinfo {author}
  {\bibfnamefont {R.}~\bibnamefont {Sakano}}, \bibinfo {author} {\bibfnamefont
  {K.}~\bibnamefont {Yoshida}}, \bibinfo {author} {\bibfnamefont
  {K.}~\bibnamefont {Shibata}}, \bibinfo {author} {\bibfnamefont
  {K.}~\bibnamefont {Hirakawa}}, \ and\ \bibinfo {author} {\bibfnamefont
  {S.}~\bibnamefont {Tarucha}},\ }\href {\doibase 10.1103/PhysRevB.81.121308}
  {\bibfield  {journal} {\bibinfo  {journal} {Phys. Rev. B}\ }\textbf {\bibinfo
  {volume} {81}},\ \bibinfo {pages} {121308} (\bibinfo {year}
  {2010}{\natexlab{b}})}\BibitemShut {NoStop}%
\bibitem [{\citenamefont {Fowler}\ and\ \citenamefont
  {Zawadowski}(1971)}]{Fowler.1971}%
  \BibitemOpen
  \bibfield  {author} {\bibinfo {author} {\bibfnamefont {M.}~\bibnamefont
  {Fowler}}\ and\ \bibinfo {author} {\bibfnamefont {A.}~\bibnamefont
  {Zawadowski}},\ }\href {\doibase
  https://doi.org/10.1016/0038-1098(71)90324-3} {\bibfield  {journal} {\bibinfo
   {journal} {Solid State Communications}\ }\textbf {\bibinfo {volume} {9}},\
  \bibinfo {pages} {471} (\bibinfo {year} {1971})}\BibitemShut {NoStop}%
\bibitem [{\citenamefont {Schlottmann}(1982)}]{Schlottmann.1982}%
  \BibitemOpen
  \bibfield  {author} {\bibinfo {author} {\bibfnamefont {P.}~\bibnamefont
  {Schlottmann}},\ }\href {\doibase 10.1103/PhysRevB.25.4815} {\bibfield
  {journal} {\bibinfo  {journal} {Phys. Rev. B}\ }\textbf {\bibinfo {volume}
  {25}},\ \bibinfo {pages} {4815} (\bibinfo {year} {1982})}\BibitemShut
  {NoStop}%
\bibitem [{\citenamefont {Bulla}\ \emph {et~al.}(2008)\citenamefont {Bulla},
  \citenamefont {Costi},\ and\ \citenamefont {Pruschke}}]{Bulla.2008}%
  \BibitemOpen
  \bibfield  {author} {\bibinfo {author} {\bibfnamefont {R.}~\bibnamefont
  {Bulla}}, \bibinfo {author} {\bibfnamefont {T.~A.}\ \bibnamefont {Costi}}, \
  and\ \bibinfo {author} {\bibfnamefont {T.}~\bibnamefont {Pruschke}},\ }\href
  {\doibase 10.1103/RevModPhys.80.395} {\bibfield  {journal} {\bibinfo
  {journal} {Rev. Mod. Phys.}\ }\textbf {\bibinfo {volume} {80}},\ \bibinfo
  {pages} {395} (\bibinfo {year} {2008})}\BibitemShut {NoStop}%
\bibitem [{\citenamefont {T\'oth}\ \emph {et~al.}(2008)\citenamefont {T\'oth},
  \citenamefont {Moca}, \citenamefont {Legeza},\ and\ \citenamefont
  {Zar\'and}}]{Toth.2008}%
  \BibitemOpen
  \bibfield  {author} {\bibinfo {author} {\bibfnamefont {A.~I.}\ \bibnamefont
  {T\'oth}}, \bibinfo {author} {\bibfnamefont {C.~P.}\ \bibnamefont {Moca}},
  \bibinfo {author} {\bibfnamefont {O.}~\bibnamefont {Legeza}}, \ and\ \bibinfo
  {author} {\bibfnamefont {G.}~\bibnamefont {Zar\'and}},\ }\href {\doibase
  10.1103/PhysRevB.78.245109} {\bibfield  {journal} {\bibinfo  {journal} {Phys.
  Rev. B}\ }\textbf {\bibinfo {volume} {78}},\ \bibinfo {pages} {245109}
  (\bibinfo {year} {2008})}\BibitemShut {NoStop}%
\bibitem [{\citenamefont {Legeza}\ \emph {et~al.}(2008)\citenamefont {Legeza},
  \citenamefont {Moca}, \citenamefont {Toth}, \citenamefont {Weymann},\ and\
  \citenamefont {Zarand}}]{Legeza.2008}%
  \BibitemOpen
  \bibfield  {author} {\bibinfo {author} {\bibfnamefont {O.}~\bibnamefont
  {Legeza}}, \bibinfo {author} {\bibfnamefont {C.~P.}\ \bibnamefont {Moca}},
  \bibinfo {author} {\bibfnamefont {A.~I.}\ \bibnamefont {Toth}}, \bibinfo
  {author} {\bibfnamefont {I.}~\bibnamefont {Weymann}}, \ and\ \bibinfo
  {author} {\bibfnamefont {G.}~\bibnamefont {Zarand}},\ }\href@noop {}
  {\enquote {\bibinfo {title} {Manual for the \uppercase{F}lexible
  \uppercase{DM-NRG} code},}\ } (\bibinfo {year} {2008}),\ \Eprint
  {http://arxiv.org/abs/0809.3143} {arXiv:0809.3143 [cond-mat.str-el]}
  \BibitemShut {NoStop}%
\bibitem [{\citenamefont {Moca}\ \emph {et~al.}(2012)\citenamefont {Moca},
  \citenamefont {Alex}, \citenamefont {von Delft},\ and\ \citenamefont
  {Zar\'and}}]{Moca.2012}%
  \BibitemOpen
  \bibfield  {author} {\bibinfo {author} {\bibfnamefont {C.~P.}\ \bibnamefont
  {Moca}}, \bibinfo {author} {\bibfnamefont {A.}~\bibnamefont {Alex}}, \bibinfo
  {author} {\bibfnamefont {J.}~\bibnamefont {von Delft}}, \ and\ \bibinfo
  {author} {\bibfnamefont {G.}~\bibnamefont {Zar\'and}},\ }\href {\doibase
  10.1103/PhysRevB.86.195128} {\bibfield  {journal} {\bibinfo  {journal} {Phys.
  Rev. B}\ }\textbf {\bibinfo {volume} {86}},\ \bibinfo {pages} {195128}
  (\bibinfo {year} {2012})}\BibitemShut {NoStop}%
\bibitem [{\citenamefont {Legeza}\ \emph {et~al.}()\citenamefont {Legeza},
  \citenamefont {Moca}, \citenamefont {Toth}, \citenamefont {Weymann},\ and\
  \citenamefont {Zarand}}]{BudapestNRG}%
  \BibitemOpen
  \bibfield  {author} {\bibinfo {author} {\bibfnamefont {O.}~\bibnamefont
  {Legeza}}, \bibinfo {author} {\bibfnamefont {C.~P.}\ \bibnamefont {Moca}},
  \bibinfo {author} {\bibfnamefont {A.~I.}\ \bibnamefont {Toth}}, \bibinfo
  {author} {\bibfnamefont {I.}~\bibnamefont {Weymann}}, \ and\ \bibinfo
  {author} {\bibfnamefont {G.}~\bibnamefont {Zarand}},\ }\href
  {http://www.phy.bme.hu/~dmnrg/} {\enquote {\bibinfo {title}
  {\uppercase{F}lexible \uppercase{DM-NRG} code,
  http://www.phy.bme.hu/\~{}dmnrg/},}\ }\BibitemShut {NoStop}%
\end{thebibliography}%

\clearpage

\section{Supplemental Information}

Here we present certain details of the perturbative  and  renormalization group calculations outlined in the main paper,
and  give some further details on  the  numerical renormalization group computations. 

\subsection{Perturbation theory}

We consider an $S_\imp ={1\over 2}$  impurity spin embedded in an s-wave BCS superconductor, with  
the impurity coupled to the spin density at position $\br = 0$. The Hamiltonian, as already   given in the main text, 
is the sum of the  Kondo interaction, 
\begin{equation}
H_{\rm K} = J \,\bS_{\rm imp} \cdot \boldsymbol{s }(0) ,
\end{equation}
and the BCS Hamiltonian
\begin{equation}
 H_{\rm host} = \sum_{\bk,\sigma}  \epsilon_\bk \,c^\dagger_{\bk\sigma} c_{\bk\sigma} + 
 \sum_{\bk,\sigma}   (\Delta \,c^\dagger_{\bk\uparrow} c^\dagger_{-\bk\downarrow} + \text{h.c.})\;.
 \end{equation}
 The spin density at $\br = 0$ is given by 
$\boldsymbol{s}(0)  = {1\over 2} \sum_{\bk, \bk',\sigma\sigma'}c^{\dagger}_{\bk\sigma}\boldsymbol{\sigma}_{\sigma\sigma'}c_{\bk'\sigma'}$, 
where  $c^{\dagger}_{\bk\sigma}$ denotes  the creation operator of electrons with momentum $\bk$ and spin $\sigma$.
The energy $\epsilon_\bk$ is measured with respect to the Fermi energy ($\epsilon_\bk = 0 \leftrightarrow E_F$), and 
we assume half filling.
We  perform a perturbative calculation in $J$ in the  free spin regime,  $T_K\ll \Delta$,
and compute the expectation value $\langle \bS^z_{\rm imp}\rangle $. In the non-interacting limit, $J=0$, the unperturbed ground state
is a direct product of the BCS ground state and the impurity spin, $\ket{\phi_0} = 
\ket{\rm{BCS}}\otimes \ket{\Uparrow}$. Here, we assume the presence of a small external magnetic field, which  lifts the spin degeneracy 
and selects  the spin up state. First order of perturbation yields a state $\ket \phi = \ket{\phi_0}+\ket {\delta \phi}$, with
\begin{equation}
\ket {\delta \phi} = -{J\over 2} \, \sum_{\sigma\sigma'}\sum_{\substack{\bk,\bk'\\ \epsilon_\bk>0\\ \epsilon_{\bk '}<0}} 
{{\bS_{\rm imp}}\over E_\bk+E_{\bk'}}\, c^\dagger_{\bk\sigma}\;{\vec \sigma}_{\sigma\sigma'}\;c_{\bk'\sigma'}\ket{\rm BCS}\otimes \ket{\Uparrow},
\nonumber
\end{equation}
with $E_\bk = \sqrt{\epsilon_\bk^2 + \Delta^2 }$ the quasiparticles' excitation energy.  
Second order corrections to the wave function can be shown to cancel and, 
to order ${\cal O}(J^2)$,  the expectation value of the impurity spin $\langle S^{z}_{\rm imp}\rangle $ 
is given by  
\begin{equation}
\langle S^{z}_{\rm imp}\rangle =\frac{\bra \phi S^{z}_{\rm imp}\ket \phi}{\bra \phi\phi\rangle}
\simeq {1\over 2}\left \{ 1 - {1\over 4}  J^2 \sum_{\bk \bk'}\frac{1}{(E_\bk+E_{\bk'})^2}\right \}.
\nonumber
\end{equation}
Replacing the momentum sums by integrals  $\sum_{\bk}\to \rho_0 \int_{-\Lambda_0}^{\Lambda_0} \rmd \epsilon $,  we obtain with logarithmic precision
\begin{equation}
\langle S^{z}_{\rm imp}\rangle = {1\over 2} \left \{1-{1\over 4} j_0^2\ln(\Lambda_0/\Delta)+ {\cal O }(j_0^3)\right \},
\end{equation}
where $j_0 = \varrho_0 J$ is the dimensionless exchange coupling. This is just  Eq.~\eqref{eq:kappa_perturbative} of the main text. 

\begin{figure}[t!]
	\includegraphics[width=0.9\columnwidth]{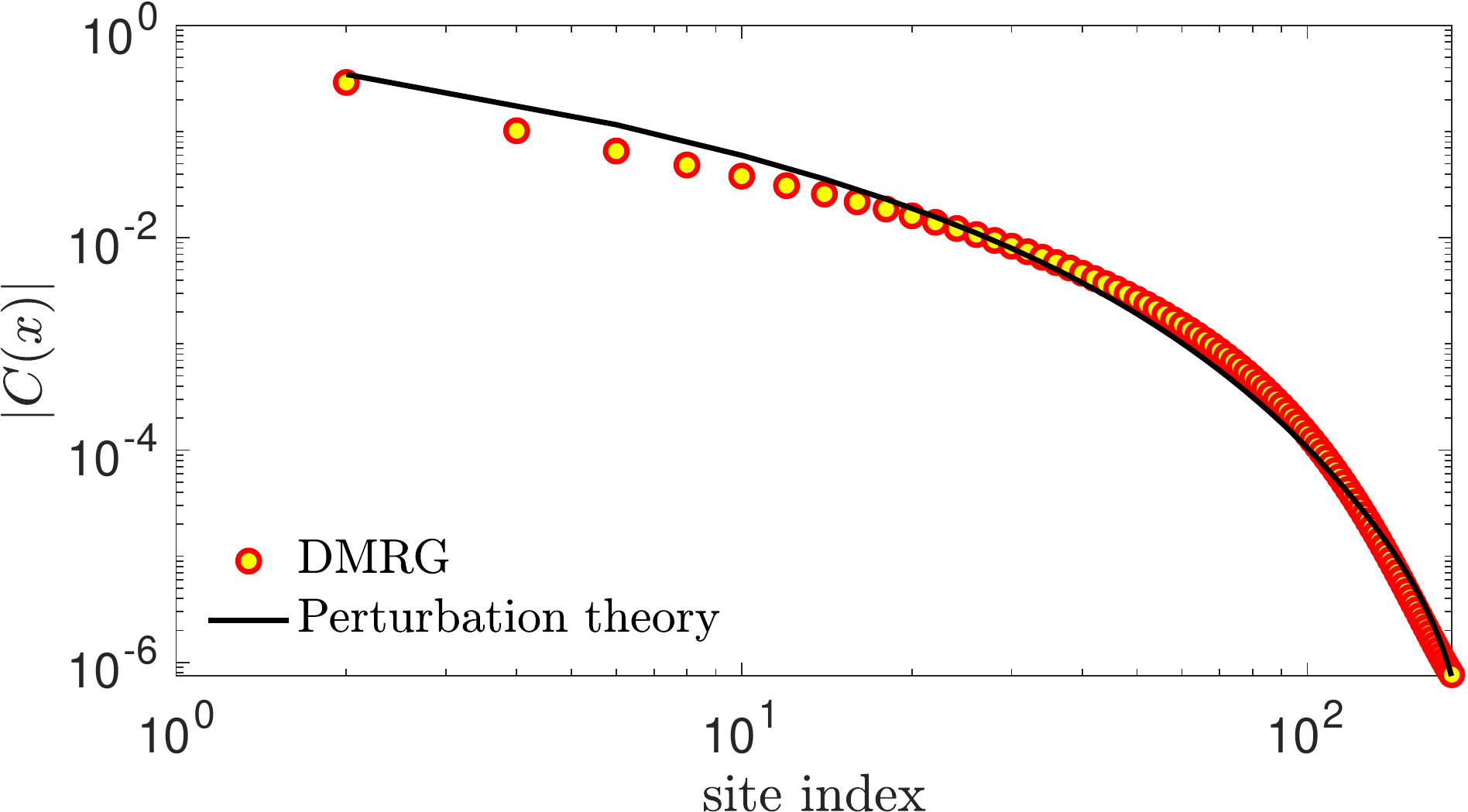}
	\caption{ Comparison of the perturbative  and  DMRG results for the envelope of spin-spin correlator in a one-dimensional superconducting chain. The system size is fixed to $L=200$ sites, $J/t = 1.8$, corresponding to $j_0 = 0.28$, and $\Delta/t = 0.1$. With these parameters $\Delta /T_K = 2.45$.}
	\label{fig:comparison}
\end{figure}

We can also use the same approach to compute the correlation $\average{\bS_\imp \cdot \bs(x)}$ in 
 a one dimensional version of the model,  where we replace $H_{\rm host}$ by 
\bea
H_{\rm chain} & = &-t \sum_{x=1}^{L-1}\sum_{\sigma} \big( c^{\dagger}_{x\,\sigma}c_{x+1\,\sigma} +{\rm h.c.}\big )
\phantom {nnn}
 \label{eq:H_chain} 
 \\
&&\phantom{n} +\; \sum_{x=1}^{L} \big(\Delta \;c^{\dagger}_{x\uparrow}c^\dagger_{x\downarrow} +\text{h.c.}\big ),  \nonumber
\eea
and couple the  impurity spin  to  the spin density at the first site, ${\bs}  = {1\over 2} \;c^{\dagger}_{1}\,{\bsigma} \,c_{1}$.   Hamiltonian~\eqref{eq:H_chain} can be solved directly in real space
by using the density matrix renormalization group (DMRG) approach.  
 Figure~\ref{fig:comparison} compares the results of  a complete DMRG 
 computation and those  of  second order
perturbation theory, which are demonstrated to  provide  good approximation for   $\average{\bS_\imp \cdot \bs(x)}$, away from the 
quantum phase transition. 

\subsection{Multiplicative renormalization group approach}

In this section we show, how one can derive Eqs.~\eqref{eq:kappa} and \eqref{eq:j(delta/T_K)} by means of the multiplicative renormalization 
group approach. The multiplicative renormalization group for the Kondo problem is best formulated in terms 
of pseudofermions, $f^\dagger_s$, used to express the spin operator  as 
$\bS_{\rm imp } = \sum_{s,s'}f^\dagger_s {\vec S}_{ss'} f_{s'}$ with the additional constraint,  
$ \sum_{s}f^\dagger_s f_s \equiv 1$. In this language, the impurity part of the Hamiltonian 
is 
\begin{equation}
H_{\rm imp} = \frac J 2 \, \sum_{\sigma,\sigma',s,s'} f^\dagger_s     \bS_{ss'} f_{s'}  
 \cdot    \psi^\dagger_\sigma\bsigma_{\sigma\sigma'}  \psi_{\sigma'} - h \sum_s s \, f^\dagger_s f_s ,
\label{eq:H_imp}
\end{equation}
with   $ \psi = \psi(0) $ the electrons' field operator at the impurity site, and the second term  a
Zeeman field, $h$, acting on the impurity spin.

Thermodynamics as well as dynamical correlations can then be 
formulated in terms of the pseudofermions' unperturbed Green's function,
${\cal G}^{(0)}_{ss'}(\tau) \equiv -i \average{{\rm T}_\tau f_s(\tau) f^\dagger_{s'}(0)}_0 $,  
the conduction electrons' unperturbed local Green's function, 
${G}_{\sigma\sigma'}^{(0)}(\tau) \equiv -i \average{{\rm T}_\tau \psi_\sigma(\tau) \psi^\dagger_{\sigma'}(0)}_0$, 
and the vertex, $\Gamma^{(0)}_{s\sigma,s'\sigma'}  = \frac J 2  \;\bS_{ss'}\cdot \bsigma_{\sigma\sigma'}$.
Similar to quantum electrodynamics,  multiplicative renormalization 
  is  a transformation
\be
\Lambda \to \Lambda', \quad J\to J', \quad h\to h', 
\ee
which  transforms the electron-impurity  vertex function,
$\Gamma$, and the impurity's  dressed Green's  function $\cal G$ multiplicatively, 
$$
{\cal G} \to Z \, {\cal G} , 
\quad 
\Gamma \to  Z^{-1}\, \Gamma \;,
$$
while it leaves the impurity contribution to the free energy, $F_\imp$ unchanged.
This transformation can be formulated in terms of simple scaling equations~\cite{Fowler.1971, Schlottmann.1982}
\bea 
\frac{\rmd j}{\rmd l} &=& j^2 - \frac 1 2 j^3 + \dots \;,
\label{eq:j}
\\
\frac{\rmd \ln h}{\rmd l} &=&  -\frac{1}{2}    j^2  + \dots  = -\frac{1}{2} \frac{\rmd j}{\rmd l} +\dots \; \;,
\label{eq:h}
\eea 
where $l = \ln(\Lambda_0/\Lambda')$ denotes the scaling variable, $j = J \varrho_0$ is the dimensionless coupling
and we have displayed terms appearing only in the 
next to leading logarithmic order. These equations are valid for   $\omega,\Lambda'\gg\Delta$, where the 
gap has only little effect and can therefore be disregarded, and must be solved 
with the initial condition, $j(\Lambda'\to \Lambda_0)=j_0$.

To compute the expectation value of the spin in the presence of a finite gap, $\Delta$, we first notice that 
the size of the spin can be obtained as 
\be 
\langle \Uparrow |S_\imp^z |\Uparrow \rangle  
= \lim_{h\to 0^+}\frac {-1}{k_B T} \frac {\partial } {\partial h}F_\imp(j, h, \Lambda_0)   \;.
\ee
However, the impurity's free energy is invariant under the renormalization group, implying that 
\be 
\frac {\partial }{\partial h}F_\imp(j, h, \Lambda_0)\;    = \frac {\partial h'}{\partial h}  \; \frac {\partial} {\partial h'} F_\imp(j', h', \Lambda')  \;, 
\ee
and therefore 
\be
\langle \Uparrow |S_\imp^z |\Uparrow \rangle_{j,\Lambda_0} =  \left(\frac {\partial h'}{\partial h}\right) \;  \langle \Uparrow |S_\imp^z |\Uparrow \rangle_{j',\Lambda'} \;.
\label{eq:spin_transformation}
\ee

If we now set the renormalized bandwidth 
equal to the superconducting gap, $\Lambda'\to \Delta$, then we have no more conduction electrons, and 
the impurity remains unscreened:
  $\langle \Uparrow |S_\imp^z |\Uparrow \rangle = 1/2$. The prefactor in 
Eq.~\eqref{eq:spin_transformation}  is thus just the $g$-factor, which we can  determine by simply integrating 
Eq.~\eqref{eq:h} to yield 
\be 
g =   \frac {\partial h'}{\partial h} = \exp \Big\{ -\frac 1 2 (j_\Delta- j_0)\Big\} \;,
\ee 
with $j_\Delta = j'(\Lambda' \to \Delta)$. This  amounts to $\kappa = 1-g$, given by Eq.~\eqref{eq:kappa}. 

To derive Eq.~\eqref{eq:j(delta/T_K)}, we integrate \eqref{eq:j} to obtain  
\be
\ln \frac {\Lambda_0} {\Lambda'} = f(j')-f(j_0)
\label{eq:renorm}
\ee
with the function $f(j)$ given as 
\be
f(j) = -\frac 1 j + \frac 1 2  \ln\,j -\frac 1 2 \ln \big(1  -\frac j 2 \big)\;.
\ee
The Kondo temperature is determined by the condition that the effective coupling be of a value $j'\equiv j^*\sim 1$, 
\be
\ln \frac {\Lambda_0} {T_K} = f(j^*)-f(j_0)\;. 
\ee
Combining this with Eq.~\eqref{eq:renorm},
we arrive at the equation, 
\be
\ln \frac {\Lambda'} {T_K} = f(j^*)-f(j')\;. 
\ee
Setting now $\Lambda'\to \Delta$ we thus obtain the implicit equation
\be 
 \frac 1 j_\Delta =  \ln \frac {\Delta} {T_K}  -  C  + \frac 1 2  \ln\,  j_\Delta +\frac {j_\Delta} 4 +\dots \;,
\ee
where $C =  f(j^*)$. An iterative solution  of this equation gives 
\be
j_\Delta\approx 
\frac 1 {\ln\left( \frac {{\cal F} \Delta} {T_K}  \right) - \frac 1 2 \ln \left(\ln\left( \frac {{\cal F} \Delta} {T_K}  \right)  \right)
+ \frac 1 {4  \ln\left( \frac {{\cal F} \Delta} {T_K}  \right) }
 }\;,
\ee
with ${\cal F} = e^{-C}$. Dropping the last, negligible  term yields the expression in the main text. 
The value of $j^*$ and thus that of $\cal F$ is somewhat arbitrary. We set it such that 
the resulting Kondo scale, $ T_K = {\cal F}\; \Lambda_0 \; \sqrt{j_0} \;\exp(-1/j_0) $ be identical to the 
Kondo scale extracted from the NRG calculations, defined there as the half-width of the 
so-called composite fermion's resonance (see next subsection). This yields the value, ${\cal F}\approx 2.5$, which allows us 
 to compare the perturbative and numerical calculations without any other adjustable parameter. 

\subsection{Details of NRG calculations}

\begin{figure}[tbh!]
	\includegraphics[width=0.9\columnwidth]{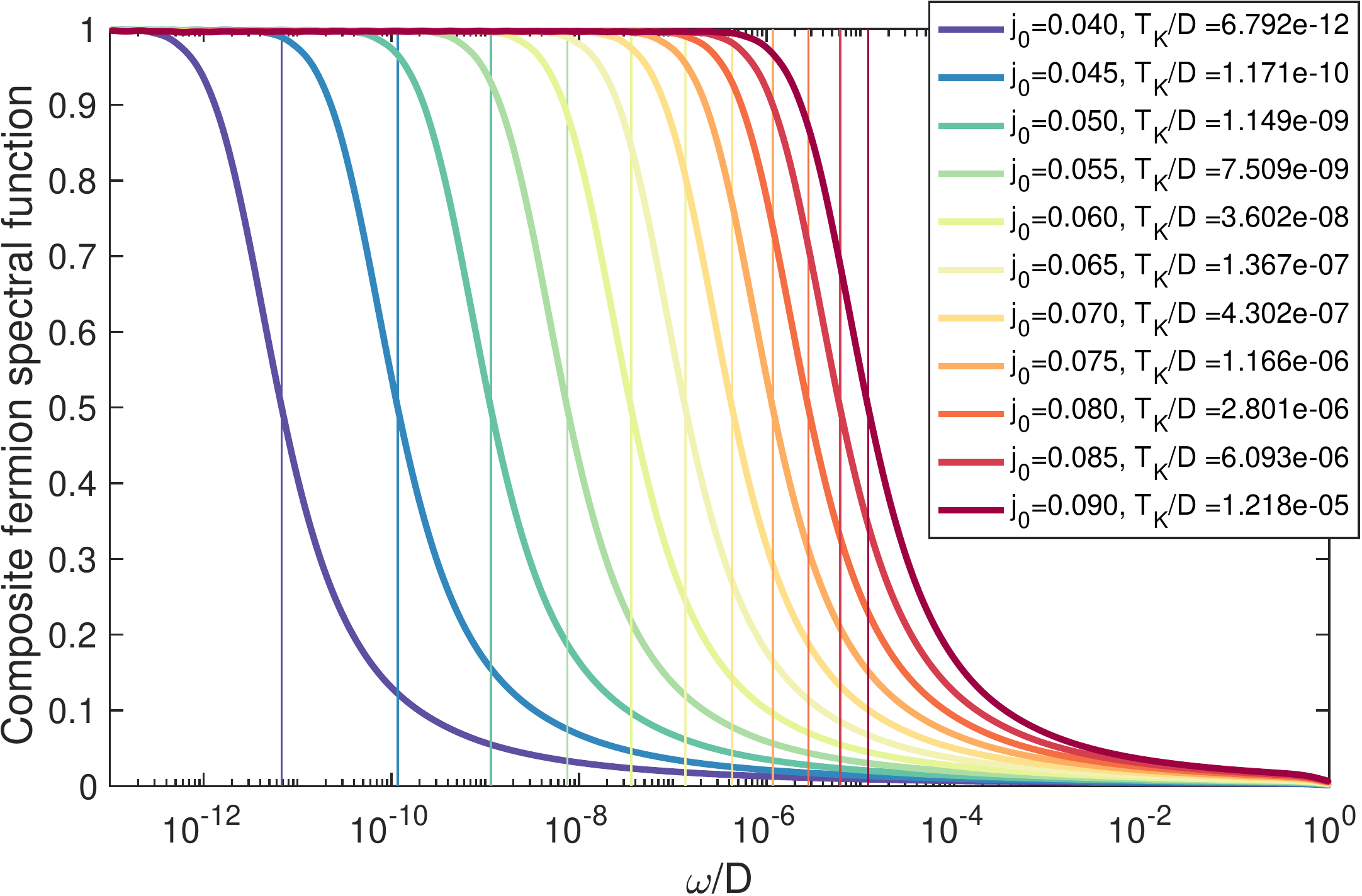}
	\caption{The spectral function for the composite fermion operator calculated for various values of the
		exchange coupling $j_0$. The Kondo temperature $T_K$ is determined
		as the half width at half maximum of the spectral function resonance.
		The calculation is done at zero temperature.}
	\label{fig:composite_fermion}
\end{figure}

In the strongly correlated regime, where the Kondo correlations are dominant,
the numerical renormalization group (NRG) method  provides accurate predictions~~\cite{Wilson.1975, Bulla.2008}. 
Contrary to DMRG, NRG works in the energy space, where it uses a logarithmic discretizition, 
allowing one  to reach  very small energy scales.
The NRG Hamiltonian defined on the Wilson chain for our problem has the form
\begin{eqnarray}
H_{\rm NRG} &=& J\; \bS_{\rm imp} \cdot {\vec s}_0
+ \sum_{i=0,\sigma}^{N} \xi_i  \big( f^{\dagger}_{i\sigma}f_{i+1\sigma} +h.c.\big ) \nonumber\\
&+& \Delta \sum_{i=0}^{N} \big( f^{\dagger}_{i\uparrow}f^\dagger_{i\downarrow} +h.c.\big ),
\label{eq:NRG}
\end{eqnarray}
where the impurity spin  is coupled by the Kondo exchange
to the local spin at site 0.  In Eq.~\eqref{eq:NRG}, 
$N$ is the length of the  chain, and $\xi_i $ denotes  hopping amplitudes,  exponentially decreasing along the chain. 
The operator  $f^{\dagger}_{i\sigma}$ denotes the creation operator at  site $i$ for a fermion with  spin $\sigma$,
and 
${\vec s}_0 = {1\over 2}\sum_{\sigma\sigma'} f^{\dagger}_{0\sigma}{\vec \sigma}_{\sigma\sigma'}f_{0\sigma'}$ is the spin density at site $i=0$. 

We solve the Hamiltonian \eqref{eq:NRG} iteratively, by keeping at least
$1024$ lowest-energy eigenstates at each step of the iteration, and
by exploiting the U(1) symmetry associated with the conservation of the total $S_z$ spin component.
For these computations, we have used  our open access flexible DM-NRG 
code~\cite{Toth.2008,Legeza.2008,Moca.2012,BudapestNRG}.

The Kondo temperature is determined as the half width at half maximum of the 
spectral function of the composite fermion, $F^\dagger = \bS_\imp \cdot \bsigma \,f^\dagger_0$.
Typical results for the composite fermion   spectral function 
are displayed in Fig.~\ref{fig:composite_fermion},
together with the corresponding values of dimensionless couplings, $j_0$, and Kondo temperatures,  $T_K$.
This comparison allows us to extract the prefactor ${\cal F}\approx 2.5 $ in Eq.~\eqref{eq:j(delta/T_K)}.  

\subsection{Details of DMRG calculations}

For the DMRG calculation we used the two-site approach introduced by White~\cite{White.1996} within the matrix product state formalism~\cite{Schollwock.2005}.
The chain Hamiltonian is given by Eq.~\eqref{eq:H_chain}, and the impurity spin is coupled to the first site. To determine the 
ground state and compute the spin-spin correlator we used the $U(1)$ symmetry for the $z$ component of the total spin $S_T^{z}$.
The chain length used in the calculations was  in general fixed to $L=200$,
but larger chain lengths, up to $L=400$ were also tested. The bond dimension $M$ was  fixed in between $400$ to $1000$. 

For each set of parameters the ground state was computed in the $S_T=0$ and $S_T=1/2$ sectors,
which allowed us to capture the parity changing transition.
Our findings for the phase diagram using DMRG match those obtained by using the NRG approach. 

\end{document}